\documentclass[aps, twocolumn, showpacs, superscriptaddress, groupedaddress]{revtex4-1}
\usepackage{array}
\usepackage{tabularx}

\usepackage{ulem}
\usepackage{graphicx}
\usepackage{yfonts}
\usepackage{amsmath}
\usepackage{amssymb}
\usepackage{amsfonts}
\usepackage{soul}
\usepackage{amsthm}
\usepackage{comment}
\usepackage{mathtools}
\usepackage{multirow}
\usepackage{dcolumn}
\usepackage{listings}
\usepackage{subfigure, rotating, bm, array}
\usepackage[utf8]{inputenc}
\usepackage[english]{babel}
\usepackage[pagebackref=false, colorlinks=true]{hyperref}
\hypersetup{
linkcolor=blue,     
citecolor=blue,     
urlcolor=blue}      

\usepackage{placeins}

\begin{document}
\title{Cosmological effect of coherent oscillation of ultralight scalar fields in a multicomponent universe}
\author{Priyanka Saha$^\dagger$, Dipanjan Dey$^\ddagger$, Kaushik Bhattacharya$^{\$}$}
\email{priyankas21@iitk.ac.in, deydipanjan7@gmail.com, kaushikb@iitk.ac.in}
\affiliation{$^\dagger, ^{\$}$ Department of Physics, Indian Institute of Technology, Kanpur,
Kanpur-208016, India\\
$^\ddagger$Beijing Institute of Mathematical Sciences and Applications, Beijing 101408, China}

\begin{abstract}
The idea that coherent oscillations of a scalar field, oscillating over a time period that is much shorter than the cosmological timescale, can exhibit cold dark matter (CDM) like behavior was previously established. In our work we first show that this equivalence between the oscillating scalar field model and the CDM sector is exact only in a flat Friedmann-Lemaitre-Robertson-Walker (FLRW) spacetime in the absence of cosmological constant and any other possible matter components in the universe when the mass of the scalar field is very large compared to the Hubble parameter. Then we show how to generalize the equivalence between the coherently oscillating scalar field model and the CDM sector in a spatially curved universe with multiple matter components. Using our general method, we will show how a coherently oscillating scalar field model can represent the CDM sector where it decays into the radiation. Our method is powerful enough to work out the dynamics of gravitational collapse in a closed FLRW spacetime where the coherently oscillating scalar field model represents the CDM sector. We have, for the first time, presented a consistent method which specifies how a coherently oscillating scalar field model, where the scalar field is ultralight, acts like the CDM sector in a multicomponent universe.   
\end{abstract}

\maketitle

\section{Introduction}

The standard model of cosmology attributes approximately 27 percent of the universe's total energy density to dark matter, an as-yet unidentified component that interacts predominantly through gravity. This estimate is supported by the final results of the full mission of the Planck satellite \cite{Planck:2018vyg}, which constrain cosmological parameters with high precision. These results are interpreted within the framework of $\Lambda$CDM cosmology, where $\Lambda$ stands for the cosmological constant and CDM specifies a cold dark matter constituent of the universe. Generally, the CDM sector is modeled by an ideal fluid with negligible pressure. There are various ways in which people have tried to model the CDM fluid, and one of the interesting ways, out of many, is related to a model of coherently oscillating scalar field as proposed by Turner in Ref.~\cite{PhysRevD.28.1243}.
Turner showed that for a quadratic potential for the scalar field, there appear to be two distinct timescales: the intrinsic oscillation period of the field related to $1/m$, where $m$ is the mass of the scalar field, and the cosmological time scale related to $1/H$,
where $H$ is the Hubble parameter. When the scalar field mass satisfies $m \gg H$, the field undergoes rapid oscillations with a period $\sim m^{-1}$, significantly shorter than the cosmological time scale $H^{-1}$. As a result, the field completes many oscillations within a single Hubble time, and its time-averaged dynamics closely resemble the dynamics of CDM in an expanding Friedmann-Lemaitre-Robertson-Walker (FLRW) spacetime. The coherently oscillating scalar field exhibits zero average pressure, and the average energy density evolves as the inverse cube of the cosmic scale-factor.

Ratra \cite{PhysRevD.44.352} later extended the analysis of Turner and produced a consistent mathematical method to deal with coherently oscillating scalar fields in cosmology. Ratra consistently worked out the background cosmological dynamics of the universe, where the scalar field was the only matter component, and then used a synchronous gauge to calculate cosmological perturbations on the background model. The scalar field dark matter model has been extensively explored in the literature. Several subsequent studies \cite{Baer:2014eja, Hidalgo:2017dfp, Hotinli:2021vxg, Garcia:2022vwm, ChoeJo:2023ffp} have explicitly adopted the approach introduced by Turner and Ratra. Hwang \cite{HWANG1997241} introduced an alternative approach by employing the uniform-curvature gauge to derive the perturbed equations, demonstrating its suitability for scalar field models. Importantly, although these studies use different gauges for spatial perturbations, the background evolution remains consistent across these approaches. The linear perturbation framework has also been applied to general power-law (scalar field) potentials, with the case $n=2$ treated as a special instance \cite{Cembranos:2015oya}. This analysis builds upon previous work and demonstrates that the resulting solutions closely resemble the perturbative behavior of dust-like perfect fluids. These methods have been further developed to include second-order perturbations \cite{Noh:2013coa} and extended to fully nonlinear analysis \cite{Noh:2017sdj}, reinforcing the result that coherently oscillating scalar fields (or axion-like fields)  exhibit cold dark matter-like behavior. Additionally, extensions of the scalar field dark matter framework have considered corrections to the standard behavior \cite{McDonald:1993ky, Cembranos:2018ulm}, including higher-order interactions or modifications to the potential. Other studies have incorporated self-interacting terms within the $\phi^{2}$ potential in the context of interacting dark sector models \cite{Aboubrahim:2024spa, vandeBruck:2022xbk}, further broadening the scope and applicability of scalar field dark matter scenarios.

Investigations into the dynamical and gravitational instabilities associated with oscillating scalar fields have also played a crucial role in assessing their cosmological viability. A detailed analysis in Ref.~\cite{Johnson:2008se} demonstrated that scalar fields with a negative equation-of-state parameter possess a large-scale dynamical instability for the growth of perturbations. This instability does not occur when the scalar field exhibits a vanishing equation-of-state parameter, making such configurations unsuitable for explaining cosmic acceleration but remaining viable as dark-matter candidates. The growth of cosmological perturbations in scalar field dark matter models has been further examined in works like \cite{Alcubierre:2015ipa, Marsh:2010wq}, confirming their relevance for structure formation scenarios. Furthermore, extensive numerical comparisons with the standard cold dark matter  paradigm, such as those in \cite{Torres:2014bpa}, have shown that scalar field and CDM perturbations evolve nearly identically in the linear regime. The primary difference lies in the presence of a characteristic cutoff in the power spectrum of the density perturbation, introduced by the scalar field dynamics. To improve the accuracy of these models, additional effort has yielded the effective fluid approximation for coherently oscillating scalar field dark matter \cite{Passaglia:2022bcr}, thereby enhancing their consistency with observational data.

From the foundational works on the equivalence of the CDM sector to the scalar field sector \cite{PhysRevD.28.1243, PhysRevD.44.352, HWANG1997241} it can be shown that, 
\begin{itemize}
\item the equivalence holds in a spatially flat FLRW spacetime. In the present work, we will show how this restriction of spatial flatness can be addressed. We will generalize the foundational results to a spatially curved FLRW spacetime.
\item In the foundational works, the equivalence only worked when the universe contained a single scalar field. In the present work, we will generalize the equivalence of the CDM sector to a scalar field sector when the universe contains various other matter fields.
\item The foundational works could not address the issue of coupling the scalar field sector to other sectors consistently, as the introduction of other sectors invariably produced a multicomponent universe. In the present work, we will show how one can introduce decaying dark matter in a formally consistent fashion. Turner \cite{PhysRevD.28.1243} did work informally with a decaying dark matter model; the present work establishes the formal way such models can be tackled.
\item In the context of interacting models, several studies have adopted the final results from the scalar field dark matter framework developed by Ratra and Turner, applying them directly to systems involving additional components or interactions \cite{HWANG20091, Mishra:2017ehw, Park:2012ru, Cardenas:2020nny, Brax:2019fzb, Urena-Lopez:2015gur}. In these works, the oscillating scalar field is treated as dark matter without explicitly deriving any conditions under which such generalizations can be done. The authors of Ref.~\cite{Mishra:2017ehw} hinted at the fine-tuning of the initial value of the ultralight scalar field. In the present work, for the first time, we will show that such generalizations require a specific constraint on the initial condition of the system.
\item Our method works for CDM models even in the end phase of radiation domination or in the dark energy-dominated phase. The calculations presented in this paper make the coherently oscillating ultralight scalar field model of CDM very general and applicable to various phases of cosmological dynamics.

\end{itemize}

We specifically focus on the mass $m = 10^{-22}$eV, called ultralight dark matter in the literature.  The reason why such an ultralight scalar field mass is chosen is related to the numerical value of the Hubble parameter which is approximately of the order of $10^{-33}$eV presently. Consequently, the value of $m$ chosen above satisfies $m \gg H$. This mass scale is particularly compelling because standard CDM models tend to over-predict small-scale structures, such as overly dense galactic cores and an excess number of dwarf galaxies, which are inconsistent with observations. Ultralight scalar particles with $m = 10^{-22}$ eV can potentially resolve these discrepancies \cite{Hu:2000ke}, as their wave-like nature suppresses small-scale gravitational clustering and prevents the formation of kiloparsec-scale cusps and substructures in dark matter halos. This suppression leads to a variety of distinctive observational signatures (see, e.g., \cite{Hui:2016ltb} and references therein). Numerous studies \cite{Suarez:2011yf, Lee:2008jp} have further examined the implications of this mass scale for cosmic structure formation. Several authors have investigated a broad spectrum of ultralight scalar field dark matter masses. A comprehensive review of axion cosmology \cite{Marsh:2015xka} highlights that a wide range of masses remains viable, depending on the specific phenomenological context. Scalar field dark matter with a mass of $10^{-23}$eV, for example, has been shown to reproduce the successes of the standard cold dark matter (CDM) model \cite{Matos:2000ss}, and this value has been adopted in subsequent studies \cite{Guzman:2003kt}. Further investigations \cite{Marsh:2013ywa} have explored an extended range of scalar field dark matter masses, typically spanning from $10^{-24}$eV to $10^{-20}$eV. The analysis includes the widely used benchmark mass of $10^{-22}$eV and indicates that masses around $10^{-21}$eV could also be viable cosmological solutions, depending on the underlying astrophysical and observational constraints. A detailed historical overview of ultralight scalar field models and their associated mass ranges can be found in \cite{Lee:2017qve}. 

The phrase \textit{mass of the scalar field} frequently appears in the literature, and it is worth clarifying its intended meaning. Typically, it refers to the mass of the quantum of the scalar field---that is, the associated particle. This identification becomes natural and direct when working in natural units, where physical quantities like mass and energy are often interchangeable. In this work, however, we adopt the geometrized unit system in which $ 8\pi G = c = 1 $, where $G$  denotes Newton’s gravitational constant and $c$ is the speed of light. Under this convention, both the scalar field mass $m$ and the Hubble parameter $H$ are expressed in units of inverse length, specifically ${\rm cm}^{-1}$. Unlike in natural units, where such quantities would naturally carry dimensions of mass or energy---our chosen system does not assign them such interpretations directly. Furthermore, we do not quantize the scalar field in this work; our analysis is entirely classical. Nevertheless, for ease of communication and consistency with standard terminology in the field, we continue to refer to $m$ as the ``mass'' of the dark matter particle---while keeping in mind that, in our unit system, it has dimensions of inverse length ${\rm cm}^{-1}$.

To demonstrate the robustness of our general approach in the presence of other cosmological components, we extend our analysis by incorporating dark energy in the form of the traditional cosmological constant $\Lambda$. It is seen that our method can incorporate the cosmological constant. The standard $\Lambda$CDM paradigm faces several theoretical and observational challenges \cite{DelPopolo:2016emo, Perivolaropoulos:2021jda}. In response to these issues, dynamical dark energy models based on scalar fields have been widely proposed. Among them, the quintessence field is one of the most commonly studied candidates. Phantom-like scalar fields, characterized by a negative kinetic term, have also been explored in modeling dark energy \cite{Bamba:2012cp, CALDWELL200223, Carroll:2003st, Copeland:2006wr, Caldwell:2003vq}. Furthermore, several recent works \cite{Saha:2024irh, Debnath:2024urb, Saha:2024xbg, Saha:2023zos} have investigated models where dynamical dark energy interacts with dark matter, highlighting the growing interest in such interacting scenarios. To exhibit the strength of our proposed solution, we have also worked with dynamical dark energy models. We show that the equivalence of the coherent oscillation of the ultralight scalar field model and the CDM sector, in presence of dynamical dark energy, can be maintained properly if we follow the rules of the equivalence as discussed in this paper.

The structure of this paper is as follows. In Section~\ref{Rapidly_oscillating_scalar_field}, we revisit the approach developed by Ratra, outlining the conditions under which a rapidly oscillating scalar field can effectively mimic CDM behavior. We also discuss the relevant averaging scheme used in this context. Section~\ref{Scalarfield_with_other_fluids} extends this framework to scenarios involving additional cosmic components, allowing for cases where the universe is not purely dust dominated. In Section~\ref{Non-minimal_coupling}, we explore how the system evolves when the scalar field is allowed to decay into radiation. Section~\ref{incs} outlines the dependence of the method on initial conditions and provides the initial values and parameter choices used consistently throughout the paper. Section~\ref{Numerical_analysis} presents a detailed numerical analysis based on appropriate initial conditions and the theoretical formulations developed. The next section \ref{gravcol} gives the results when our method is extended for models involving gravitational collapse. Finally, we summarize our results and provide concluding remarks in the last section.
\section{The traditional model: Only one rapidly oscillating scalar field and CDM-like features}
\label{Rapidly_oscillating_scalar_field}

In this section, we will point out why the traditional equivalence of the coherently oscillating scalar field model and the CDM sector only holds in a spatially flat FLRW spacetime in the presence of only one kind of matter: the scalar field.
We will initially follow the exposition by Ratra Ref.~\cite{PhysRevD.44.352}, whose work first gave a concrete mathematical realization of the equivalence in terms of oscillating fields. The basic calculations in the aforementioned paper indirectly show why the method fails when we have a multicomponent universe.

We assume the maximally symmetric FLRW spacetime with the metric 
\begin{eqnarray}
ds^2=-dt^2 + a^2(t)\left[\frac{dr^2}{1-kr^2} + r^2 (d\theta^2 + \sin^2\theta d\phi^2 )\right].
\end{eqnarray}
The curvature parameter $k$, taking values $0$, $+1$, and $-1$, distinguishes between flat, closed, and open FLRW spacetimes. Here $a(t)$ is the scale factor, which is dimensionless when $k = 0$, making the comoving coordinate $r$ carry the dimension of length. For $k = \pm 1$, $a(t)$ has the dimension of length and $r$ becomes dimensionless. In this section, we restrict our analysis to the spatially flat case with $k = 0$.

If we have a single scalar field with the harmonic potential 
\begin{eqnarray}
 V(\phi)=\frac12 m^2 \phi^2\,,    
\end{eqnarray} 
where $m$ is the scalar field mass, the background scalar field equation is
\begin{eqnarray}
\ddot{\phi}  + 3 H\dot{\phi} + m^2\phi=0\,,
\label{canscbt}
\end{eqnarray}
where $H=\dot{a}/a$ is the Hubble parameter. We assume 
\begin{eqnarray}
m \gg H\,,    
\label{bmhc}
\end{eqnarray}
so that there are two widely separated time scales (in natural units)\footnote{To compare our results with others we will often use the natural units but the calculations in our paper use geometrized units.} in the problem: one related to $1/m$ and may be interpreted as the time period of oscillations of the scalar field in the harmonic potential, and the other time scale is the cosmological scale related to $1/H$. To proceed from this point, we will specifically write the background field in terms of two different functions of $t$, one of which changes slowly with $t$, (changes in a time scale much greater than $1/m$), and another one which changes faster with respect to $t$ (changes in a time scale $1/m$). Following \cite{PhysRevD.44.352} we can write:
\begin{eqnarray}
\phi(t)=\phi_+(t)\sin \alpha(t) + \phi_-(t)\cos \alpha(t)\,,
\label{phi0exp}
\end{eqnarray}
where $\alpha(t)$ changes rapidly in a time scale $1/m$ , whereas $\phi_{\pm}(t)$ changes much slowly compared to the rate of change of $\alpha(t)$. Here $\phi_\pm(t)$ has scalar field dimension, whereas $\alpha$ is dimensionless. We can now differentiate the above function as:
$$\dot{\phi}=\dot{\phi}_+\sin \alpha + \dot{\phi}_-\cos \alpha + \dot{\alpha}(\phi_+\cos \alpha - \phi_-\sin \alpha)\,,$$
and a subsequent differentiation gives
\begin{eqnarray}
\ddot{\phi} &=&\ddot{\phi}_+\sin \alpha + \ddot{\phi}_-\cos \alpha + \dot{\alpha}(\dot{\phi}_+\cos \alpha - \dot{\phi}_-\sin \alpha)
\nonumber\\
&+&\ddot{\alpha}(\phi_+\cos \alpha - \phi_-\sin \alpha) +\dot{\alpha}(\dot{\phi}_+\cos \alpha - \dot{\phi}_-\sin \alpha)\nonumber\\ &-& \dot{\alpha}^2 \phi\,.
\label{}
\end{eqnarray}
Using these expressions of the derivatives in the scalar field equation, we see that there are two terms which are proportional to $\phi$ on the left side of the field equation. We can group them and demand that this term vanish. This gives the condition:
\begin{eqnarray}
\dot{\alpha}^2 - m^2=0\,.
\label{psieqn}
\end{eqnarray}
This equation can be solved for $\alpha$; we have $\dot{\alpha} = m$ (taking the positive square root), whose solution is $\alpha(t)=m(t-t_i)$ where we can set $t_i=0$.

The other nonzero terms in the field equation are:
\begin{eqnarray}
&&\ddot{\phi}_+\sin \alpha + \ddot{\phi}_-\cos \alpha + \dot{\alpha}(\dot{\phi}_+\cos \alpha - \dot{\phi}_-\sin \alpha)\nonumber\\
&&+ \ddot{\alpha}(\phi_+\cos \alpha - \phi_-\sin \alpha)+ \dot{\alpha}(\dot{\phi}_+\cos \alpha - \dot{\phi}_-\sin \alpha)\nonumber\\
&&+ 3H\left[\dot{\phi}_+\sin \alpha + \dot{\phi}_-\cos \alpha\right.\left.  + \dot{\alpha}(\phi_+\cos \alpha - \phi_-\sin \alpha)\right]\nonumber\\&&=0\,.
\nonumber
\end{eqnarray}
Substituting the forms of $\dot{\alpha}$ and $\ddot{\alpha}=0$ in the above equation, we get:
\begin{eqnarray}
&&\ddot{\phi}_+\sin \alpha + \ddot{\phi}_-\cos \alpha + 2m(\dot{\phi}_+\cos \alpha - \dot{\phi}_-\sin \alpha)\nonumber\\
&&+ 3H\left[\dot{\phi}_+\sin \alpha + \dot{\phi}_-\cos \alpha
 + m(\phi_+\cos \alpha - \phi_-\sin \alpha)\right]\nonumber\\&&=0\,.
\label{mordered}
\end{eqnarray}
We will solve the above equation for all the orders of the mass term $m$. This turns out to be a valid and interesting solution.
Grouping all the terms which are proportional to $m$, and setting them to zero, gives:
\begin{eqnarray}
&&2(\dot{\phi}_+\cos \alpha - \dot{\phi}_-\sin \alpha)
+ 3H(\phi_+\cos \alpha - \phi_-\sin \alpha)\nonumber\\&&=0\,.
\label{mordered1}
\end{eqnarray}
The above equation can be grouped in terms of $\cos\alpha$ and $\sin\alpha$. We have:
\begin{eqnarray}
\left(2\dot{\phi}_+ + 3H\phi_+\right)\cos \alpha - \left(2\dot{\phi}_-   
+ 3H\phi_-\right)\sin \alpha=0\,.
\label{mordered2}
\end{eqnarray}
In the above equation, the functions $\phi_\pm$, $\dot{\phi}_\pm$, and $H$ vary in the cosmological timescale, whereas the functions $\cos \alpha(t)$, $\sin \alpha(t)$ vary much faster than the aforementioned functions. For this reason, we can assume the terms multiplying  $\cos \alpha(t)$ and $\sin \alpha(t)$ in the above equation to be separately equal to zero. This requirement yields:
\begin{eqnarray}
\dot{\phi}_{\pm} + \frac32 H\phi_{\pm} =0\,.
\label{mordered3}
\end{eqnarray}
Ultimately demanding the other terms in Eq.~(\ref{mordered}) (terms proportional to $m^0$) vanish, we get:
$$\ddot{\phi}_+\sin \alpha + \ddot{\phi}_-\cos \alpha + 3H(\dot{\phi}_+ \sin \alpha + \dot{\phi}_- \cos \alpha)=0\,.$$
This equation yields:
\begin{eqnarray}
\ddot{\phi}_{\pm} + 3 H\dot{\phi}_{\pm}=0\,.
\label{mordered4}
\end{eqnarray}
If the scalar field solution can consistently be represented in the form given in Eq.~(\ref{phi0exp}), then both Eq.~(\ref{mordered3}) and Eq.~(\ref{mordered4}) have to be followed simultaneously. We will see that both of these equations will not be simultaneously followed in a flat FLRW spacetime when there are any other matter components in the universe\footnote{It can be verified that for $a(t)\propto t^{\frac23}$ both the above equations i.e., Eq.~(\ref{mordered3}) and Eq.~(\ref{mordered4}) are simultaneously satisfied.}.
The solutions of Eq.~(\ref{mordered3}) are of the form:
\begin{eqnarray}
\phi_\pm(t)=\frac{C_\pm}{a(t)^{3/2}}\,,
\label{phitime}
\end{eqnarray}
where $C_\pm$ are constants and have the same dimension as that of a scalar field. 

Till now, we have solved the scalar field equation in a spatially flat FLRW background in terms of the yet unknown scale-factor $a(t)$. The solution of the sale-factor comes from the Einstein equations. The solution of the Friedmann equations will ultimately predict whether the above scheme of solution for the scalar field is consistent. As the scalar field is supposed to be oscillating frequently on a cosmological time scale we average out the rapid fluctuations in the scalar field sector and write one of the Friedmann equations as:
\begin{eqnarray}
  \left(\frac{\dot{a}}{a}\right)^2 =&& \frac{1}{3}\langle \rho \rangle =\frac{1}{3}\left\langle \frac{\dot{\phi}^2}{2}  + \frac12 m^2 \phi^2\right\rangle\nonumber\\&& = \frac{1}{6}\langle \dot{\phi}^2  +  m^2 \phi^2\rangle\,,
\label{dfried1}  
\end{eqnarray}
where the time component of the energy momentum tensor gives, $-T_0^0=\langle \rho \rangle$\,.
Here, the angular brackets specify the time average over a time period $2\pi/m$. It must be noted that this averaging is over the oscillating time scale of the scalar field, not a complete time average over cosmological time. As a consequence, this time average of various quantities will not eradicate all time dependence of the quantities. 

\begin{table*}[t!]
\centering
\begin{ruledtabular}
\begin{tabular}{|l|l|l|l|l|l|}
\multicolumn{6}{|c|}{Comparison of Oscillating Scalar Field CDM Models} \\
\hline
Model type &
One scalar field &
Spatial curvature &
Multiple components &
Interacting CDM &
Special Initial condition \\
\hline
Ratra--Turner type &
Yes &
No &
No &
No &
Insensitive \\
\hline
SDB type &
Yes &
Yes &
Yes &
Yes &
Sensitive \\
\end{tabular}
\end{ruledtabular}
\caption{Table showing the comparison of our work with the foundational works on this area. Here ``SDB'' stands for the last names of authors of the present paper, Saha-Dey-Bhattacharya.}
\label{tab1}
\end{table*}

The average of the product of two functions of time, as $\langle s(t)f(t) \rangle$, where $s(t)$ varies much slower in time compared to one period of oscillation for the fast oscillating function, $f(t)$ can be written as $\langle s(t)f(t) \rangle =s(t)\langle f(t) \rangle$. If $f(t)$ has a time period $T=2\pi/m$ then
\begin{eqnarray}
\langle f(t) \rangle = \frac{m}{2\pi}\int_0^{\frac{2\pi}{m}} f(t^\prime)\,dt^\prime\,.
\label{averg}
\end{eqnarray}
Using this averaging scheme we can show that the average energy density of the system is given as:
\begin{eqnarray}
\langle\rho\rangle=\frac12 m^2(C_+^2 + C_-^2)\frac{1}{a^3}\left(1+ \frac98 \frac{H^2}{m^2}\right)\,, 
\label{rhsfried14}
\end{eqnarray}
and the average pressure is:
\begin{eqnarray}
  \langle P \rangle =\frac{9}{16}m^2 (C_+^2 + C_-^2)\frac{1}{a^3}\left(\frac{H^2}{m^2}\right)\,.
\label{pressb}  
\end{eqnarray}
The details of the calculation is given in appendix \ref{appa}. The ratio of the pressure and the energy density of the system then yields:
\begin{eqnarray}
\frac{\langle P \rangle}{\langle\rho\rangle}= \frac{\frac98\frac{H^2}{m^2}}{\left(1+ \frac98 \frac{H^2}{m^2}\right)}\,.
\label{beos}
\end{eqnarray}
In the limit $m \gg H$, we see that the above ratio tends to zero, specifying an effective equation of state (EoS) of dust. To get the functional form of the scale-factor one has to use the above average values in the first Friedmann equation, Eq.~(\ref{dfried1}). Doing so one obtains:
\begin{eqnarray}
a(t)=a_{i}\left[1 + A(t-t_i)\right]^{2/3}\,,
\label{asoln}
\end{eqnarray}
where $a_{i}$ is the scale-factor at $t=t_{i}$ and
\begin{eqnarray}
A^2 \equiv \frac{3m^2}{8a_i^3}(C_+^2 + C_-^2)\,.
\label{Adef}
\end{eqnarray}
This is the matter-dominated universe solution. It is shown in appendix \ref{appa} that when $H^2\gg m^2$ and $\langle P \rangle \ll \langle \rho \rangle$, the other Friedmann equation (the acceleration equation) is also satisfied. 
The crucial point about this solution is that only the above matter-dominated solution of the scale-factor satisfies Eq.~(\ref{mordered4}). It is seen from the above calculations when there is only one coherently oscillating scalar field in the universe, oscillating in a time scale which is much shorter than the cosmological time scale, the cosmological system behaves like a dark matter dominated system and only in these case both Eq.~(\ref{mordered3}) and Eq.~(\ref{mordered4}) are simultaneously satisfied. If the scale-factor is different from the form of the scale-factor in pure matter domination, then Eq.~(\ref{mordered4}) will never be satisfied and the whole calculational scheme will become inconsistent.

In the next section we will show how this model can be generalized for a multicomponent universe which can be even spatially curved.  Our generalized model for oscillating ultralight scalar fields can even accommodate interaction with other components of the universe. 
The Ratra-Turner model is insensitive to initial condition as the equivalence holds only when the scalar field solution is given by
the expression in Eq.~(\ref{phitime}) and we do not have to specify some specific initial condition for the scalar field. We do not obtain Eq.~(\ref{phitime}) for some particular initial condition. The generalization of the Ratra-Turner like model requires a particular kind of initial condition of the problem.  This topic will be discussed in the next section. A brief comparison of the various oscillating ultralight scalar field models, producing CDM like features, is presented in Table \ref{tab1}.  

\section{Inclusion of other matter components and the generalization of the coherently oscillating scalar field solution}
\label{Scalarfield_with_other_fluids}

The equivalence of the scalar field system and the CDM like sector presented in the last section is only an exact equivalence in spatially flat FLRW spacetime, where the only matter component in the universe is the scalar field, as only in that case, both 
Eq.~(\ref{mordered3}) and  Eq.~(\ref{mordered4}) are simultaneously satisfied. In the presence of any other matter component or spatial curvature energy, the equivalence does not hold. On the other hand, the averaging techniques introduced in the previous section are very useful.  We will attempt to generalize the previous averaging techniques in a multicomponent universe. In the general case, we still assume the scalar field potential to be harmonic, as given in the last section. As long as we have $m \gg H$ there can be coherent oscillations in the scalar field sector, and the time period of these oscillations can be much smaller than the cosmological time scale. To keep the averaging techniques introduced in the last section, in the general model, we will have to veer out of the scheme presented in the last section at some point and establish a new dynamical system. We do so in this section. In the previous section, there were two assumptions:
\begin{enumerate}

\item The separation of mass or time scales, given by $m \gg H$.

\item The energy density $\rho$ is solely coming from a dust solution.  
\end{enumerate}
In the present case, we would like to retain the first assumption, but we will violate the second assumption. 

In the present case, we assume the field can be written as in Eq.~(\ref{phi0exp}) and proceed similarly as we did in the last section till Eq.~(\ref{mordered}): 
\begin{eqnarray}
&&\ddot{\phi}_+\sin \alpha + \ddot{\phi}_-\cos \alpha + 2m(\dot{\phi}_+\cos \alpha - \dot{\phi}_-\sin \alpha)\nonumber\\
&&+ 3H\left[\dot{\phi}_+\sin \alpha + \dot{\phi}_-\cos \alpha
 + m(\phi_+\cos \alpha - \phi_-\sin \alpha)\right]\nonumber\\&&=0\,,
\nonumber
\end{eqnarray}
and separate out the $\sin\alpha$ and $\cos\alpha$ terms and write the equation as:
\begin{eqnarray}
&&[\ddot{\phi}_+ -2m\dot{\phi}_- +3H(\dot{\phi}_+ - m \phi_-)]\sin \alpha \nonumber\\&&+ [\ddot{\phi}_- + 2m\dot{\phi}_+ + 3H( \dot{\phi}_-
 + m\phi_+)]\cos\alpha=0\,.\nonumber\\
\label{morderedn}
\end{eqnarray}
We have very slowly varying functions multiplying $\sin\alpha$ and $\cos\alpha$  functions which vary with time much rapidly. As discussed before, in such cases, the slowly varying functions separately have to vanish so that the above equation holds. As a consequence, we have:
\begin{eqnarray}
\ddot{\phi}_+ -2m\dot{\phi}_- +3H(\dot{\phi}_+ - m \phi_-)&=&0\,,
\label{gen1}\\
\ddot{\phi}_- + 2m\dot{\phi}_+ + 3H( \dot{\phi}_- + m\phi_+) &=& 0\,.
\label{gen2}
\end{eqnarray}
If we again segregate the terms in the above equation depending upon the powers of $m$ then we will get back Eq.~(\ref{mordered3}) and Eq.~(\ref{mordered4}), which will only produce a purely dust dominated universe. This point was explained in the first paragraph of this section.
Consequently, in the presence of other matter sources, we do not simplify the above solutions and work with Eq.~(\ref{gen1}) and Eq.~(\ref{gen2}) as a set of coupled differential equations in $\phi_\pm$. Here we assume that the universe is not purely dust dominated, although $m \gg H$. In the presence of another perfect fluid component and in the presence of spatial curvature, the Friedmann equations are:
\begin{eqnarray}
  \left(\frac{\dot{a}}{a}\right)^2 &=& \frac{1}{3}\left[\langle \rho \rangle + \tilde{\rho} + \Lambda\right]-\frac{k}{a^2}\,,
\label{cfried1}\\
\frac{\ddot{a}}{a}&=&-\frac{1}{6}\left[\langle \rho + 3P \rangle + (\tilde{\rho} + 3\tilde{P})\right]\,,
\label{cfried2}
\end{eqnarray}
where $\tilde{\rho}, \tilde{P}$ arise due to another barotropic fluid or due to the presence of another scalar field and $\Lambda$ specifies the cosmological constant. 
\subsection{Inclusion of spatial curvature and dealing with a multicomponent universe}

Here $k=0$ or $k=\pm 1$ specifies the uniform curvature of the spatial 3-dimensional hypersurfaces in the FLRW spacetime. In the previous section, we did not have any spatial curvature energy density as it was purely one component system, and there was only one form of energy density that was present. While we generalize the previous ideas to multicomponent fluids, we can include the curvature energy density as it represents just another kind of energy density which can contribute to the dynamics. If the extra matter sector is coming from another barotropic fluid, then we must have an extra equation:
\begin{eqnarray}
\dot{\tilde{\rho}} + 3H (\tilde{\rho} + \tilde{P})=0\,.
\label{extraem1}
\end{eqnarray}
The average energy density of the oscillating scalar field is obtained from Eq.~(\ref{rhsfried1}):
\begin{eqnarray}
\langle\rho\rangle &=&
\frac14 (\dot{\phi}_+^2 + \dot{\phi}_-^2) + \frac{m^2}{2}({\phi}_+^2 + {\phi}_-^2) \nonumber\\
&& +\frac{m}{2}(\dot{\phi}_-\phi_+ - \dot{\phi}_+ \phi_-) \,. 
\label{rhsfried14n}
\end{eqnarray}
One must note that in the present case, although we are using the same notation to write the functions $\phi_\pm(t)$, they are unknown functions. The average pressure of the oscillating scalar field is given by 
\begin{eqnarray}
  \langle P \rangle = \frac14  (\dot{\phi}_+^2 + \dot{\phi}_-^2) + \frac{m}{2}(\dot{\phi}_-\phi_+ - \dot{\phi}_+ \phi_-)\,.
\label{pressbn}  
\end{eqnarray}
We can assume the other fluid has an equation of state (EoS) as:
\begin{eqnarray}
\tilde{P}=\tilde{\omega} \tilde{\rho}\,.
\label{neos}
\end{eqnarray}
We have four unknowns: $\phi_+(t), \phi_-(t), a(t), \tilde{\rho}(t)$ now. The initial conditions must satisfy Eq.~(\ref{cfried1}).
One can always calculate the EoS in the scalar field sector as:
\begin{eqnarray}
&&\omega = \frac{\langle P \rangle}{\langle \rho \rangle}\nonumber\\&&=
\frac{\frac14  (\dot{\phi}_+^2 + \dot{\phi}_-^2) + \frac{m}{2}(\dot{\phi}_-\phi_+ - \dot{\phi}_+ \phi_-)}{\frac14 (\dot{\phi}_+^2 + \dot{\phi}_-^2) + \frac{m^2}{2}({\phi}_+^2 + {\phi}_-^2)  + \frac{m}{2}(\dot{\phi}_-\phi_+ - \dot{\phi}_+ \phi_-)}\,,\nonumber\\
\label{eosds}
\end{eqnarray}
and see how it evolves with time. This EoS in principle will be a function of time. If the barotropic fluid is absent, then $\omega \to 0$, but in the presence of the fluid one may have a different EoS in the scalar field sector. Later, we will see that if we implement the proper initial conditions, then the equivalence between the coherently oscillating ultralight scalar field sector and the CDM sector remains valid for all times, and as a consequence, this $\omega \sim 0$. The other fluid can represent radiation or the dark energy, where the barotropic fluids have $\tilde{\omega}=1/3$ or $\tilde{\omega}=-1$.

When there are multiple components of matter in the universe, it is sometimes useful to understand how the effective EoS varies with time. We define effective EoS, $\omega_{\rm eff}$ as:
\begin{eqnarray}
\omega_{\rm eff} = \frac{\langle \rho \rangle + \tilde{\rho}}{\langle P \rangle + \tilde{P}}\,.
\label{effeos}
\end{eqnarray}
The behavior of the effective EoS tells us the nature of the effective matter component of the universe.

\subsection{Introducing dynamic dark energy}

As was stated before, we will sometimes assume $\tilde{\rho}, \tilde{P}$ arising from another scalar field sector. The other scalar field will be assumed to be representing the dynamic dark energy sector. In those cases, one will work purely with the dynamic dark energy sector and set $\Lambda=0$ in Eq.~(\ref{cfried1}). This scalar field can represent either quintessence
($\epsilon=+1$) or phantom ($\epsilon=-1$) like fields, in the presence of a potential $V(\psi)=V_{m} e^{-\lambda \psi}$, where the scalar field equation is:
\begin{eqnarray}
\epsilon \ddot{\psi}+3H\epsilon\dot{\psi} + \frac{dV}{d\psi}=0\,.
\label{kGeqforpsi}
\end{eqnarray}
Here $V_{m}$ and $\lambda$ are parameters which specify the dynamic dark energy potential. This form of potential for dynamic dark energy is widely used in the literature \cite{Ratra:1987rm, Ferreira:1997hj, Sahni:2002kh, Copeland:2006wr, Tsujikawa:2013fta}. The corresponding energy density and pressure are:
\begin{eqnarray}
\tilde{\rho} &=& \rho_{\psi} = \frac{\epsilon\dot{\psi}^2}{2} + V_{m}e^{-\lambda\psi}\,,
\label{rhopsi}\\
\tilde{P}&=&P_{\psi} = \frac{\epsilon\dot{\psi}^2}{2} - V_{m}e^{-\lambda\psi}\,. 
\label{ppsi}
\end{eqnarray}
When we are mainly interested in observing the effect of dynamical dark energy in the background expanding universe, which contains a CDM component and is assumed to be spatially flat, we will set $k=0$ in Eq.~(\ref{cfried1}). On the other hand, if we want to see how the dark sector behaves in a gravitational collapse of a slightly overdense spherical patch, we will assume $k=1$ in  Eq.~(\ref{cfried1}).  Next, we discuss how one can include decaying of the CDM sector into the radiation.

\subsection{Deacy of the CDM sector and radiation}\label{Non-minimal_coupling}

In Ref.~\cite{PhysRevD.28.1243} Turner extended the theory of coherently oscillating scalar fields to the case where this scalar field sector decays into radiation, where the scalar field undergoing damped oscillations in the harmonic potential may dump energy into the radiation sector. The oscillations are damped due to the friction produced by the expansion of the universe. In Ref.~\cite{PhysRevD.28.1243}, the effect of this decay was dealt with in a heuristic manner, as there was no attempt to observe the backreaction effect of the produced radiation field on the oscillation phenomenology.  In our method, we can exactly take into account the effect of the CDM sector decay into the radiation sector. 

In the present case, we will have $\tilde{\rho}=\rho_{R}$ and $\tilde{P}=P_{R}$ where $\rho_{R}, P_{R}$ are the radiation energy density and radiation pressure. Instead of starting with Eq.~(\ref{canscbt}), we now start with the modified scalar field equation as:
\begin{eqnarray}
\ddot{\phi}  + (3 H + \Gamma)\dot{\phi} + m^2\phi=0\,.
\label{canscbtn}
\end{eqnarray}
where $\Gamma$ gives the decay rate of the scalar field. This decaying dark matter scheme is appealing because, in general, axion-like particles have a small electromagnetic coupling \cite{Raffelt:1987im} and hence axions can decay to photons.
Suppose the scalar field decays into radiation, then, Eq.~(\ref{extraem1}) gets modified to:
\begin{eqnarray}
\dot{\rho}_{R} + 4H \rho_{R} = \Gamma\dot {\phi}^2\,, 
\label{mextraem1}
\end{eqnarray}
giving us the rate at which radiation energy density builds up. Applying the averaging scheme, we have
\begin{eqnarray}
 &&\dot{\rho}_{R} + 4H \rho_{R} = \Gamma \langle \dot{\phi}^2 \rangle\nonumber\\&&= \frac{\Gamma}{2} \left[\dot{\phi}_+^2 + \dot{\phi}_-^2 + m^2({\phi}_+^2 + {\phi}_-^2) \right] \nonumber\\&&+ \Gamma m(\dot{\phi}_-\phi_+ - \dot{\phi}_+ \phi_-)\,,
 \label{radend}
\end{eqnarray}
where we have used Eq.~(\ref{phi0d2}). One can verify that Eq.~(\ref{gen1}) and Eq.~(\ref{gen2}) now become:
\begin{eqnarray}
\ddot{\phi}_+ -2m\dot{\phi}_- +(3H+\Gamma)(\dot{\phi}_+ - m \phi_-)&=&0\,,
\label{gen1n}\\
\ddot{\phi}_- + 2m\dot{\phi}_+ + (3H+\Gamma)( \dot{\phi}_- + m\phi_+) &=& 0\,.
\label{gen2n}
\end{eqnarray}
In general, $\Gamma$ can be a constant or a function of $\phi$. To show the feasibility of our method, in this paper, we have only worked with a constant $\Gamma$. If there is no radiation considered in the system, then the system starts from a point where $\rho_{R}=0$ and the radiation density builds up. In the case of constant, $\Gamma$  we can have the weak dissipative regime where $\Gamma < 3H$  or the strong dissipative regime where $\Gamma > 3H$. Because $\Gamma$ is the rate at which dark matter sector decays into the radiation sector, we work in the weak dissipative regime. For the equivalence between the coherently oscillating scalar field sector and the CDM sector, one must be careful about the initial point and ensure $m \gg 3H + \Gamma$.

We will see how these various cases can be solved using the method discussed in this section. Before we proceed to tackle definite models of multicomponent cosmological evolution, we first introduce the initial conditions in the present scenario. The dynamics developed here crucially depend upon a particular class of initial conditions, as discussed in the next subsection.
\subsection{Intricacy of the initial conditions}
\label{incs}

In this paper, we are considering a manifold $(\mathcal{M},g_{\alpha\beta})$ that is globally hyperbolic, i.e., there exists a global Cauchy surface ($\Sigma$) whose domain of dependence $D(\Sigma)\equiv \mathcal{M}$. From the Cauchy data given on a Cauchy surface $(\Sigma)$, we can construct the entire manifold. In the present case the initial data, which involves the scalar field and its derivative at the initial time, $t=0$, must have to be specified on $\Sigma$. A brief discussion on the formalism of the Cauchy initial value problem is presented in appendix \ref{formal}.   

The discussions in the last section established the equivalence between the oscillating scalar field model and the CDM sector, and one may for all practical purpose forget about the scalar field sector and assume that the universe consists of a barotropic fluid whose energy density falls as $a(t)^{-3}$ and whose pressure is zero. On the other hand, in a multicomponent universe, one has to test whether such an equivalence still persists, as in this case, it is difficult to give an analytic proof of the equivalence between the scalar field sector and the CDM sector. Even if one assumes that the previous equivalence holds, then also in the presence of non-minimal coupling of the scalar field $\phi$ with another field or fluid, or in case of decay of scalar field into another fluid or field, one has to solve for the field $\phi$. We will first see that even in the simplest model of the equivalence, as discussed in the previous section, the initial constraints on $(a(0),~ H(0),~\phi_{\pm}(0),~\dot{\phi}_{\pm}(0))$ which lead to the equivalence are nontrivial. To show the intricacies involved, for the time being, we assume that $\tilde{\rho}=\tilde{P}=0$, $k=0$, $\Lambda=0$, and we have a universe purely dominated by the scalar field energy density. We would like to solve the problem discussed in the last section using the equations derived in this section. In such a case, the Friedmann equations are given by Eq.~(\ref{dfried1}) and Eq.~(\ref{secfried}), where $\langle \rho \rangle$ and $\langle P \rangle$ are given in Eq.~(\ref{rhsfried14n}) and Eq.~(\ref{pressbn}). 
The initial data $(a(0),~ H(0),~\phi_{\pm}(0),~\dot{\phi}_{\pm}(0))$ that satisfy the Friedmann constraint in Eq.~(\ref{cfried1}) is not enough to reproduce the exact equivalence in the limit $m \gg H$ we obtained in the last section. We observe that the initial values of $\phi_\pm(0)$ must satisfy the condition in Eq.~(\ref{initcons}) on $\Sigma$. If this additional constraint is not satisfied with our initial choice of the functions, then we will never get the previous exact results, as those results were obtained when $\phi_\pm(0)$ and $\dot{\phi}_\pm(0)$  satisfied the aforementioned condition. If we take into account the condition given in Eq.~(\ref{phitime}) and take the time derivative of $\phi_\pm(t)$ (arising from it) then we see that the initial conditions should be such that
\begin{eqnarray}
\frac{\dot{\phi}_+(0)}{\phi_+(0)} = \frac{\dot{\phi}_-(0)}{\phi_-(0)} = -\frac32 H(0)\,. 
\label{initcn}
\end{eqnarray}
If the initial data $(a(0),~ H(0),~\phi_{\pm}(0),~\dot{\phi}_{\pm}(0))$ are chosen to satisfy the Friedmann constraint along with the additional constraints mentioned above, then it can be shown that the scalar field component closely mimics the behavior of cold dark matter (CDM), even in the most general case where multiple matter sources are present in the universe and the scalar field is non-minimally coupled or decaying into to other components.

\begin{table*}[t]
\begin{ruledtabular}
\begin{tabular}{|l|l|l|l|l|}
\multicolumn{5}{|c|}{Conditions of validity for the solution} \\
\hline
$m \gg H$ &
$k = 0,\pm1$ &
$\Gamma < 3H$ &
$m \gg 3H + \Gamma$ &
Initial condition \\
\hline
\parbox{2.6cm}{Fundamental constraint, ensuring separation of time scales} &
\parbox{2.6cm}{Valid for all $k$} &
\parbox{2.6cm}{Valid for constant weak dissipation of dark matter } &
\parbox{3.2cm}{Additionally required for separation of time scales in a dissipative system } &
\parbox{2.6cm}{Eq.~(\ref{initcn}) must be satisfied} \\
\end{tabular}
\end{ruledtabular}
\caption{Conditions and constraints under which the solution is valid.}
\label{tab2}
\end{table*}

In the present context, it would be important to investigate the specific type of evolution of $(a(t), \phi(t))$ for which the above constraint (Eq.~(\ref{initcn})) propagates along time. Let's consider 
\begin{eqnarray}
    \mathcal{E_+}(t) &=&\dot{\phi}_+(t)+\frac32~H(t)\phi_+(t),\\
    \mathcal{E_-}(t) &=&\dot{\phi}_-(t)+\frac32~H(t)\phi_-(t).
\end{eqnarray}
For these constraints, one can check that they remain zero, i.e., $\mathcal{E_\pm}(t)=0$, for all times when we impose the initial constraint as given in Eq.~(\ref{initcn}) and the scale-factor solution is $a(t)\propto (1+ At)^{2/3}$ . For this initial constraint, one can have a particular solution of the scale-factor, as given above, for which the terms proportional to $m$ in
Eq.~(\ref{gen1}) and Eq.~(\ref{gen2}) segregate out, and individually (those terms separately) become zero (as it happened in the previous section).  These facts suggest that when we have $\mathcal{E_\pm}(0)=0$ initially on $\Sigma$, we can have the dust-like solution and the constraints propagate in time. If $a(t)$ has any other time dependence, one may not have $\mathcal{E_\pm}(t) = 0$ on $\Sigma$ and in general $\mathcal{E_\pm}(t)$ will evolve with time\footnote{More analytical work is needed to fully reveal the dynamics of the constraint in general. We hope to work out the dynamics of the constraint propagation in a future publication.}.  Therefore, if $\mathcal{E}_{\pm}|_{t=0}=0$, then $\mathcal{E}_{\pm}(t)=0$ for all time $t$, i.e., the additional constraint Eq.~(\ref{initcn}) on the Cauchy surface $\Sigma$ would propagate along time when we have a pure dust dominated universe. If other non-dust-like matter is present along with dust, then the scale-factor $a(t)$ would not have the similar form, and therefore the additional constraint Eq.~(\ref{initcn}) at the Cauchy surface on the scalar fields would not propagate exactly.

In our model, we consider a scalar field with mass $m \gg H$, which introduces two distinct time scales. These time scale differences can have interesting effects in Eq.~(\ref{gen1}) and Eq.~(\ref{gen2}).  Denoting the shorter time scale by $\tau=1/m$,
we can write these equations as:
\begin{eqnarray}
\left(\tau\frac{d}{dt}\dot{\phi}_++3H\tau\frac{d}{dt}\phi_+\right) -2\left(\dot{\phi}_- +\frac32H \phi_-\right)&=&0\,,
\label{genM1}\\
\left(\tau\frac{d}{dt}\dot{\phi}_-+3H\tau\frac{d}{dt}\phi_-\right) +2\left(\dot{\phi}_+ +\frac32H \phi_+\right)&=&0.
\label{genM2}
\end{eqnarray}
In the above equations, $\tau d/dt$ of any function  $\dot{\phi}_\pm$ or $\phi_\pm$ simply specifies the change of these functions in the time interval $\tau$. We have assumed that these functions change negligibly in the time interval $\tau$, and consequently, the terms in the brackets on the left side of the (left hand side of the) above equations (which contain $\tau$) are negligible, and consequently, the other terms (in the above equations) must also be negligible. In Ref.~\cite{PhysRevD.44.352}, the above equations become solvable because all the terms in the brackets, appearing in the above equations, individually vanish for all times $a(t)\propto (1 + At)^{2/3}$. In the present case, all the bracketed terms in the above equations remain nonzero; they have negligible values because $m\gg H$. The initial conditions play a special role because we have exactly set $\left(\dot{\phi}_{\pm}(0) +\frac32 H(0)~\phi_{\pm}(0)\right) = 0$ at the Cauchy surface, hence initially all the bracketed terms in the above equations are zero. Because $a(t) \ne a_i(1+ A t)^{2/3}$ in our case, the bracketed terms will not remain exactly zero for later times but will pick up negligible values as we have maintained $m \gg H$ for all times. The constraint in the initial condition, as given in Eq.~(\ref{initcn}), makes the above equations consistent. If we have not assumed $\left(\dot{\phi}_{\pm}(0) +\frac32 H(0)~\phi_{\pm}(0)\right) \sim 0$ initially, then the initial conditions should have contradicted  Eq.~(\ref{gen1}) and Eq.~(\ref{gen2}) when $m\gg H$. We can say that when $a(t)\propto (1 + At)^{\frac23}$ the constraint $\left(\dot{\phi}_{\pm}(0) +\frac32 H(0)~\phi_{\pm}(0)\right)= 0$ exactly propagates along time, whereas in a multicomponent universe, the constraint propagates approximately. This approximation works very well when $m \gg H$. 
 
From the above discussion we can now conclude that if we allow $k \ne 0$ and more matter components and the cosmological constant in the universe where the Friedmann equations are given by Eq.~(\ref{cfried1}) and Eq.~(\ref{cfried2}) and the quantities $\phi_\pm(0), \dot{\phi}_\pm(0), a(0), \dot{a}(0)$ satisfy the conditions given in Eq.~(\ref{initcn}) then the equivalence between the oscillating scalar field sector and the CDM sector remains valid as long as $m\gg H$.
Even if the scalar field $\phi$ decays into some other component, the initial conditions in Eq.~(\ref{initcn}) and the condition $m\gg H$ play an active role in preserving the aforementioned equivalence.

In the present case the initial conditions on the oscillating scalar field, in the general case, are chosen in such a way that when all the other matter components of the universe are removed from the system, the dynamics of the system settles down to the one component system where the equivalence between the coherently oscillating scalar field model and CDM sector holds perfectly for $m\gg H$.  The statements about the equivalence of the coherent oscillating scalar field model and the CDM sector, in the most general case, depend upon the choice of the initial conditions stated here. The statements made in this section will be numerically verified in various cases, and we will show that the generalization of the aforementioned equivalence can in principle, be done for all practical cases in cosmology.   

Before we specify the numerical estimate of the initial values, we summarize the conditions and constraints of our solution in Table \ref{tab2}. The table shows the basic requirements and constraints of the method we have proposed in this article. 
\subsection{Numerical estimation of the initial values in geometrized units}

Before validating our claims, we will specify some important points about the geometrized units used in our work and explain how they influence the choice of initial conditions. In this unit system, quantities such as length, mass, and time share the same dimensionality, expressed in inverse centimeters ($\text{cm}^{-1}$), while the scalar field \(\phi\) remains dimensionless. For broader relevance and comparison, we refer to \textit{natural units}, where energy is commonly expressed in electron volts (eV). In natural units, we consider the mass range of the oscillating scalar field $\phi$ to be $m\sim 10^{-22}~\text{eV}$, which is considered an ultralight scalar field, while the present-day Hubble constant is $H\sim 10^{-33}~\text{eV}$. In geometrized units, we adopt a scalar field mass of $m=1.93 \times 10^{-17}~\text{cm}^{-1}$.  Using the conversion factor $1~\text{cm}^{-1}=0.19733\times 10^{-4}~\text{eV}$ our chosen $m$ value corresponds to $m=3.8\times 10^{-22}~\text{eV}$ in natural units, consistent with ultralight scalar field dark matter models.

We further adopt the cosmological parameters from the final results of the \textit{Planck} satellite full mission, \cite{Planck:2018vyg} which provide the following present-day values of the density parameters of dark energy ($\Lambda$), cold dark matter (M), baryonic matter (B) and radiation (R) as: $\Omega_{\Lambda_{0}}=0.68$, $\quad \Omega_{\text{M}_{0}}h^{2}=0.120,\quad \Omega_{\text{B}_{0}}h^{2}=0.0224,\quad\Omega_{\text{R}_{0}}h^{2}=4.15 \times 10^{-5}$ with the dimensionless Hubble constant given by $h=0.67$. The redshift z is related to the scale-factor $a$ by the standard relation: $\text{z} = \frac{a_{0}}{a} - 1$, with current scale-factor $a_{0}=1$. The scale-factor at radiation-matter (cold dark matter + baryonic matter) equality is given by, $a_{\text{eq}}=\frac{\Omega_{\text{R}_{0}}}{\Omega_{\text{M}_{0}}+\Omega_{\text{B}_{0}}}=2.8\times 10^{-4}$ corresponding to a redshift of $\text{z}_{\text{eq}}=3571$. Similarly, the scale-factor at matter-dark energy equality is, $\tilde{a}_{\text{eq}}=\frac{\Omega_{\text{M}_{0}}+\Omega_{\text{B}_{0}}}{\Omega_{\Lambda_{0}}}=0.77$ corresponds to a redshift of $\tilde{\text{z}}_{\text{eq}}=0.3$. The subscript "$0$" denotes the values at the current epoch of the universe.

The dark energy component in this study is modeled using two different approaches. The first approach employs the cosmological constant $\Lambda$, with a fixed value of $\Lambda=1.08\times 10^{-56}~\text{cm}^{-2}$, derived from the observed dark energy density parameter, and in geometrized unit the vacuum energy density $\rho_{\Lambda}=\Lambda$. The second approach introduces a dynamical scalar field $\psi$, which can be a quintessence or a phantom field. In both cases, the scalar field is governed by an exponential potential of the form $V(\psi)=V_{m}e^{-\lambda \psi}$, where the parameter $\lambda=1$, is fixed throughout the work for both field types. The maximum potential value $V_{m}$ is chosen such that the density parameter of the scalar field matches that of the cosmological constant at the present epoch. We set $V_{m}=1.38\times10^{-56}~\text{cm}^{-2}$ for the quintessence field and $V_{m}=0.91\times10^{-56}~\text{cm}^{-2}$ for the phantom field.

To handle extremely small numerical values efficiently, we define a reference unit $u = 10^{-28}~\text{cm}^{-1}$, which corresponds to $10^{-33}~\text{eV}$ in natural units. All parameters are expressed in terms of $u$ throughout this paper, thus the value of $m=1.93\times 10^{11}~u$, $V_{m}=1.38~u^{2}, 0.91~u^{2}$ respectively for quintessence and phantom field and $\rho_{\Lambda}=\Lambda = 1.08~u^2 $. We require two initial conditions to solve the scalar field equation Eq. (\ref{kGeqforpsi}). These are chosen as $\dot{\psi}(t_{i})=0.0001~u$ and $\psi(t_{i})=0.0001$ at the initial time $t_{i}$ and are held fixed throughout the analysis. The initial time $t_{i}$ and the initial value of the scale factor $a=a_{i}$ depend on the cosmological epoch under consideration. The scale factor becomes dimensionless when a flat universe $k=0$ is assumed, and in cases where the curvature constant  $k=1$,  the scale factor carries the dimension of length. The specific values used for $a_{i}$ will be clarified in the relevant sections.

The energy density $\rho(t)$ for any component is related to the density parameter using $\Omega(t) = \rho(t)/\rho_{c}(t)$, where the critical density in geometrized units is given by $\rho_{c}(t) = 3H(t)^2$. In the case of an oscillating scalar field, as we are using time-averaged energy density, the corresponding density parameter is defined as $\Omega_{\phi}(t) =\langle \rho(t)\rangle/\rho_{c}(t)$.  At the present epoch, the critical density takes the value $\rho_{c_{0}}=1.60~u^{2}$. Using this, the present-day dark matter density parameter is expressed as, $\Omega_{\text{M}_{0}}=\frac{\rho_{\text{M}_{0}}}{a^{3}_{0}\rho_{c_{0}}}$, and similarly, for baryonic matter and radiation, we have, $\Omega_{\text{B}_{0}}=\frac{\rho_{\text{B}_{0}}}{a^{3}_{0}\rho_{c_{0}}}, \quad \Omega_{\text{R}_{0}}=\frac{\rho_{\text{R}_{0}}}{a^{4}_{0}\rho_{c_{0}}}$. Using these relations, and $a_{0}=1$ we obtain $\rho_{\text{M}_{0}}=0.42~u^{2},\quad \rho_{\text{B}_{0}}=0.079~u^{2},\quad \rho_{\text{R}_{0}}=1.4\times 10^{-4}~u^{2}.$   All these choices, as specified in this section, along with the initial conditions and parameter values, are held fixed throughout our calculations unless stated otherwise.

\begin{figure*}
\subfigure[Evolution of the time-averaged EoS parameter ($\omega$) of dark matter with time, where the system starts in the Matter-Domination era.]
{\includegraphics[width=80mm,height=50mm]{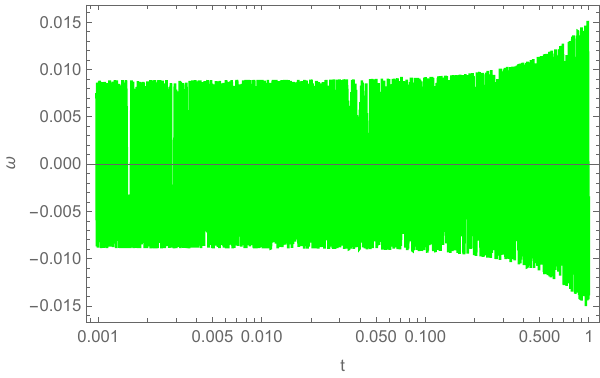}\label{matterdominationw}}
\hspace{0.2cm}
\subfigure[Evolution of effective EoS parameter $\omega_{\text{eff}} = (\langle \rho \rangle + \tilde{\rho}) / (\langle P \rangle + \tilde{P})$ with time, where the system starts in the Matter-Domination era. Here, dark matter quantities are time-averaged ($\langle \cdot \rangle$), while dark energy quantities are not.
]
{\includegraphics[width=80mm,height=50mm]{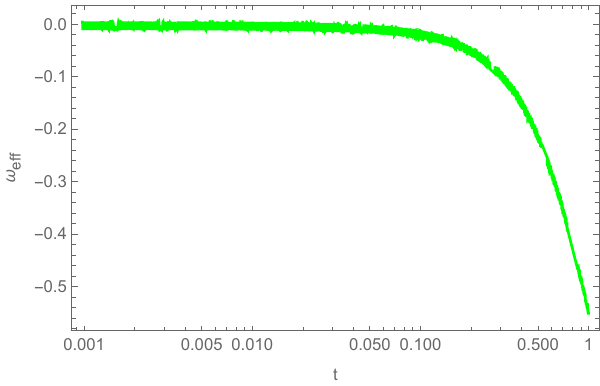}\label{matterdominationweff}}
\hspace{0.2cm}
\subfigure[Evolution of the time-averaged EoS parameter ($\omega$) of dark matter with time, where the system starts in the Radiation-Domination era.]
{\includegraphics[width=80mm,height=50mm]{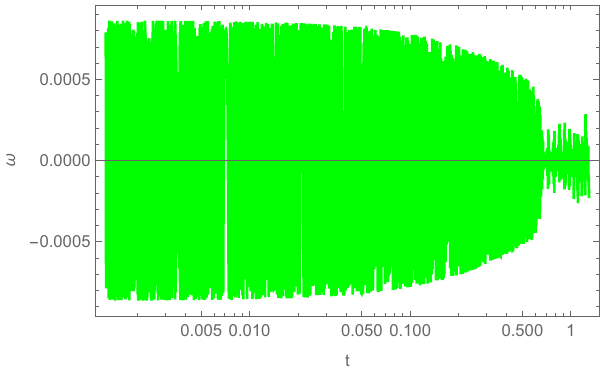}\label{radiationdominationw}}
\hspace{0.2cm}
\subfigure[Evolution of effective EoS parameter $\omega_{\text{eff}} = (\langle \rho \rangle + \tilde{\rho}) / (\langle P \rangle + \tilde{P})$ with time, where the system starts in the Radiation-Domination era. Here, dark matter quantities are time-averaged ($\langle \cdot \rangle$), while dark energy quantities are not.]
{\includegraphics[width=80mm,height=50mm]{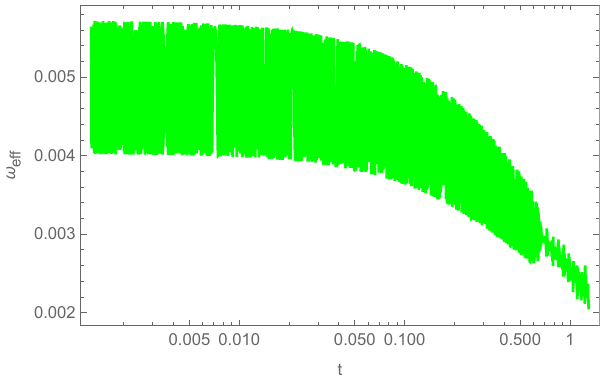}\label{radiationdominationweff}}
\hspace{0.2cm}
\subfigure[Evolution of the time-averaged EoS parameter ($\omega$) of dark matter with time, where the system starts in Dark energy-Domination era.]
{\includegraphics[width=80mm,height=50mm]{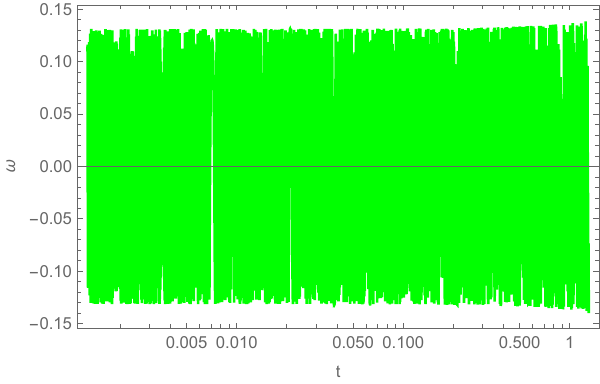}\label{darkenergydominationw}}
\hspace{0.2cm}
\subfigure[Evolution of effective EoS parameter $\omega_{\text{eff}} = (\langle \rho \rangle + \tilde{\rho}) / (\langle P \rangle + \tilde{P})$ with time, where the system starts in Dark energy-Domination era. Here, dark matter quantities are time-averaged ($\langle \cdot \rangle$), while dark energy quantities are not.]
{\includegraphics[width=80mm,height=50mm]{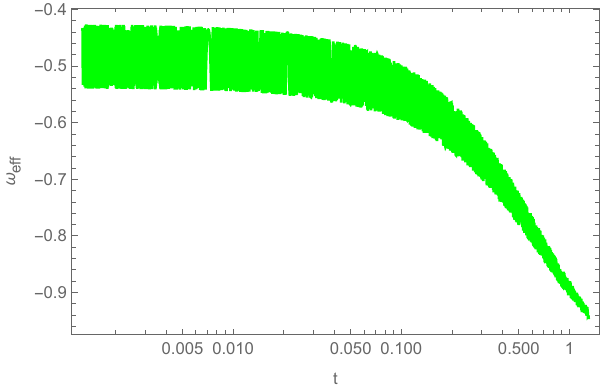}\label{darkenergydominationweff}}
\hspace{0.2cm}
\caption{\footnotesize Evolution of the effective and dark matter equation-of-state (EoS) parameters with respect to cosmic time (t) across different cosmic regimes, where dark matter is modeled as an oscillating scalar field $\phi$ and dark energy is represented by a cosmological constant with $\Lambda = 1.08\,u^{2}$. The initial conditions deviate from those specified in Eq.~(\ref{initcn}). The initial conditions correspond to distinct redshifts: $z = 3600$ for the radiation-dominated era, $z = 5$ for the matter-dominated era, and $z = 0.2$ for the dark energy-dominated era. The mass of the dark matter scalar field is fixed at $m = 1.93\times10^{11}\,u$, and the initial conditions for the scalar fields satisfy the relation $\frac{\dot{\phi}_+(t_i)}{\phi_+(t_i)} = \frac{\dot{\phi}_-(t_i)}{\phi_-(t_i)} = -\tfrac{3}{2} \times 10^{3} H(t_i)$ with the initial dark matter energy densities chosen according to the corresponding redshifts for each era. All other cosmological parameters follow the Planck 2018 results~\cite{Planck:2018vyg}, and all quantities are expressed in geometrized units, $u = 10^{-28}\,\mathrm{cm}^{-1}$.}

\label{differentinitialconditions}
\end{figure*}

\subsection{Deviation from initial conditions}

To a great extent, it is true that the averaging process goes on properly, for various background cosmological evolutions, as long as the oscillating frequency of the scalar field is larger than the universe expansion rate, but the result becomes more and more exact only when proper initial conditions are applied. The point can be explained by an example. Suppose the universe contains both radiation and the oscillating scalar field, where the oscillating scalar field represents the dark matter sector. A natural requirement of such a system is that if we initially set the energy density of radiation to be zero, then the system evolves in a purely matter-dominated phase, where the scalar field satisfies all the equations set in the last section. This can only happen if we set the initial condition as given in Eq.~(\ref{initcn}). This initial condition naturally takes care of the pressure-less fluid model discussed in the last section; any other initial condition will not reproduce those results accurately. This fact shows that even if $H/m \ll 1$ and whatever be the background, Eq.~(\ref{initcn}), ensures that the oscillating scalar field model works properly. In the presence of other components, of course, the conditions on the cosmological evolution change and affect the scalar field evolution, but the effect on the scalar sector is small as shown by Eq.~(\ref{genM1})  and Eq.~(\ref{genM2}). As one deviates from the initial condition in Eq.~(\ref{initcn}), the averaging procedure starts to get disrupted. Random initial conditions on $\phi_\pm$ and $\dot{\phi}_\pm$ do not produce any interesting result, but modification of the initial condition in Eq.~(\ref{initcn}) in a particular way produces results which are themselves oscillatory, as we will show in this subsection. For strong deviations from the initial conditions, the averaging procedure itself gives oscillatory solutions.

If the system is already in the matter-dominated regime, then the theory in the last section is valid and no additional assumptions are needed. Our method is applicable in all regimes, and therefore, a consistent initial condition becomes essential for obtaining reliable results in all regimes. To demonstrate this, we have shown below the evolution of the equation-of-state (EOS) parameter for both the cold dark matter component and the total EOS parameter across different regimes. These results highlight how the system behaves when the initial condition proposed in our work, specifically the relation $\frac{\dot{\phi}_+(t_{i})}{\phi_+(t_{i})} = \frac{\dot{\phi}_{-}(t_{i})}{\phi_-(t_{i})} = -\frac32 H(t_{i})\,$ given in Eq. (\ref{initcn}) is not satisfied, thereby demonstrating the importance of choosing appropriate initial conditions for a consistent treatment.

The results in Fig.~[\ref{differentinitialconditions}] are shown for three different regimes: radiation domination, matter domination, and dark energy domination. In each case, the initial conditions were chosen such that they do not satisfy the relation in Eq. (\ref{initcn}), and this leads to noticeable deviations from the expected cold dark matter behavior in our method. It is important to note that in all cases, the condition $m\gg H$ is satisfied.

For the matter domination era, the energy densities are evaluated at redshift $z=5$, which corresponds to a scale factor of $a_{\text{i}}=1.6\times10^{-1}$. At this initial scale factor, the radiation energy density is given by $\rho_{\text{R}_{i}}=\rho_{\text{R}_{0}}/a_{i}^{4}=0.21~u^{2}$, the baryonic matter density is $\rho_{\text{B}_{i}}=\rho_{\text{B}_{0}}/a_{i}^{3}=19~u^{2}$, and the cold dark matter energy density is $\rho_{\text{M}_{i}}=\rho_{\text{M}_{0}}/a_{i}^{3}=102~u^{2}$. Similarly, for the radiation domination and dark energy domination, the background energy densities are evaluated at the appropriate redshifts. Where the radiation-dominated epoch has been considered at $ z = 3600 $, and the dark energy-dominated epoch at $z = 0.2$. Using the mass of the oscillating scalar field, the appropriate initial condition is, $\frac{\dot{\phi}_+(t_{i})}{\phi_+(t_{i})} = \frac{\dot{\phi}_-(t_{i})}{\phi_-(t_{i})} = -\frac32 H(t_{i})$, but in this case, we intentionally deviate from the required initial condition and instead choose the ratio as, $-\frac{3}{2}\times10^{3} H(t_{i})$ to investigate the sensitivity of the system to the initial setup. The resulting evolution of the EoS parameter for the scalar field and the total EoS for different regimes are shown in Fig.~[\ref{differentinitialconditions}]. The deviation from the standard initial condition is relatively large, as smaller deviations will not lead to appreciable changes in the output. This amount of deviation from the standard result depends upon the values of the parameters of the model, in particular, the mass of the scalar field. As the energy density will be fixed at some redshift value, a larger mass for the scalar field will give smaller values of $\phi$ and $\dot{\phi}$, and a small change in them may not appreciably change the result. As one requires an appreciable change of the initial condition, where the change is of the above type, the previous authors working in this topic may not have noticed the specific initial conditions required to solve the system in a smooth manner.

\section{Verification and applicability of our method in the spatially flat expanding FLRW spacetime}
\label{Numerical_analysis}

In this section, we will first verify the predictions made in the previous sections at different levels of complexity. This verification serves as proof of our proposal of the general equivalence of the coherently oscillating ultralight scalar field model and the CDM sector in the general case where there are multiple matter components and the system satisfies the condition $m \gg H$. 
Initially, we will show that the equivalence works perfectly when we have three more components in the universe, these components being radiation, baryonic matter, and dark energy. Next, we will increase the level of complexity and include the case where CDM sector decays into radiation in the presence of dark energy and baryonic matter. The latter case distinctly shows the applicability of the solution presented in this paper in more complicated cosmological settings.
\begin{figure*}
\subfigure[Evolution of density parameters ($\Omega$) with redshift (z) in a cosmological model with dark energy modeled as the cosmological constant ($\Lambda$).]
{\includegraphics[width=80mm,height=50mm]{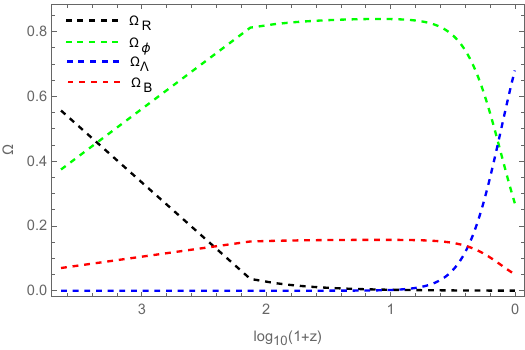}\label{radiation_matter_darkenergy1}}
\hspace{0.2cm}
\subfigure[Evolution of the density parameters ($\Omega$) with redshift (z) in a cosmological model with dark energy modeled as a quintessence scalar field ($\psi$).
]
{\includegraphics[width=80mm,height=50mm]{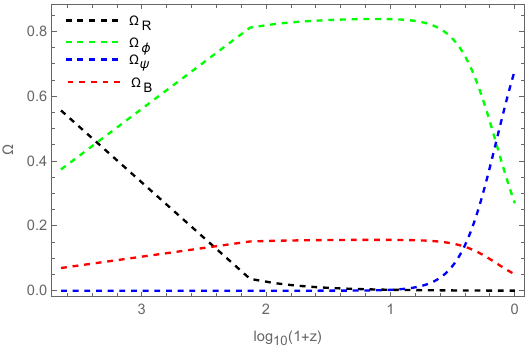}\label{radiation_matter_darkenergy2}}
\hspace{0.2cm}
\subfigure[Evolution of density parameters ($\Omega$) with redshift (z) in a cosmological model with dark energy modeled as a phantom scalar field ($\psi$).]
{\includegraphics[width=80mm,height=50mm]{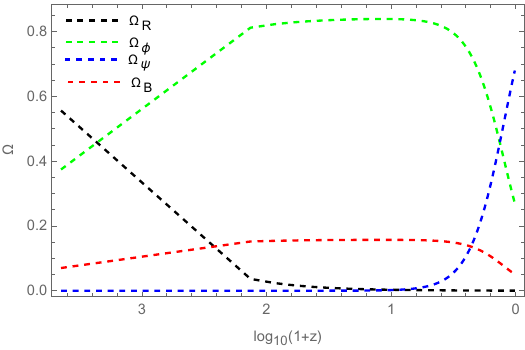}\label{radiation_matter_darkenergy3}}
\hspace{0.2cm}
\subfigure[Evolution of time-averaged dark matter equation of state (EoS) parameter ($\omega$) with redshift (z) for different dark energy models.]
{\includegraphics[width=80mm,height=50mm]{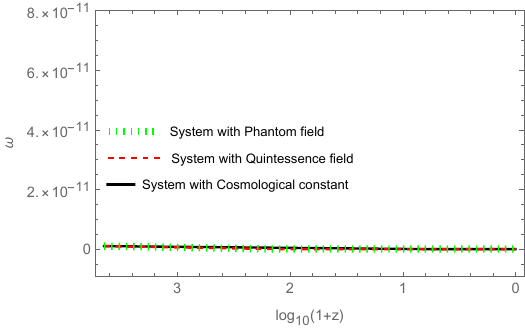}\label{radiation_matter_darkenergy4}}
\hspace{0.2cm}
\subfigure[Evolution of dark energy equation of state (EoS) parameter ($\omega_{\text{DE}}$) with redshift (z) for different dark energy models.]
{\includegraphics[width=80mm,height=50mm]{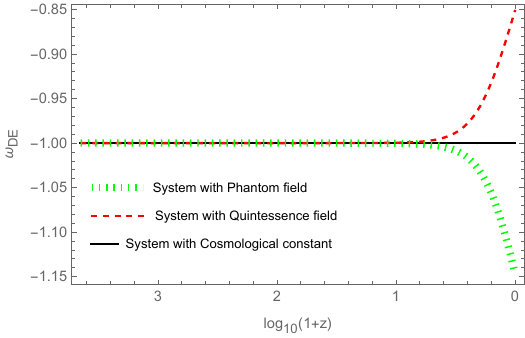}\label{radiation_matter_darkenergy5}}
\hspace{0.2cm}
\subfigure[Evolution of effective equation-of-state (EoS) parameter $\omega_{\text{eff}} = (\langle \rho \rangle + \tilde{\rho}) / (\langle P \rangle + \tilde{P})$ with redshift for different dark energy models. Dark matter quantities are time-averaged ($\langle \cdot \rangle$), while dark energy quantities are not.]
{\includegraphics[width=80mm,height=50mm]{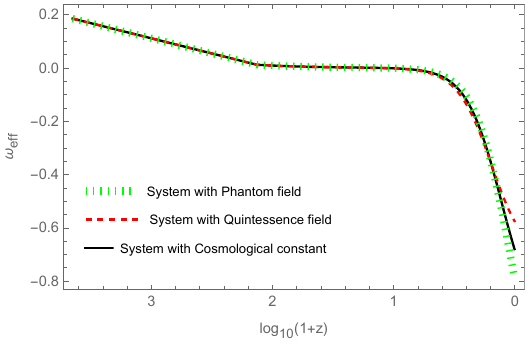}\label{radiation_matter_darkenergy6}}
\hspace{0.2cm}
\renewcommand{\figurename}{\footnotesize Figure}
\caption{\footnotesize Evolution of cosmological variables as a function of redshift (z) in a universe composed of radiation (R), baryons (B), dark matter ($\phi$), and dark energy. The horizontal axis shows $\log_{10}(1+z)$, plotted in decreasing order to represent the forward progression of cosmic time. Dark energy is modeled in three ways: (1) a cosmological constant with $\Lambda = 1.08\, u^2$; (2) a quintessence scalar field ($\psi$) with exponential potential $V(\psi) = V_m e^{-\lambda \psi}$, $V_m = 1.38\, u^2$, $\lambda=1$, and initial conditions $\psi(t_i) = 0.0001$, $\dot{\psi}(t_i) = 0.0001~u$; (3) a phantom scalar field with the same potential form and same initial and parameter values except $V_m = 0.91\, u^2$. Dark matter is an oscillating scalar field $\phi$ with mass $m = 1.93\times 10^{11}\, u$ and initial conditions $\phi_\pm(t_i) = 10^{-6}$, $\dot{\phi}_\pm(t_i) = -\frac{3H_i}{2} \times 10^{-6}~u$, with $a_i = 2.24\times10^{-4}$ and $H_i = 1.82\times10^5~u$. All other cosmological parameters follow Planck 2018 \cite{Planck:2018vyg}, and quantities are expressed in geometrized units $u = 10^{-28}~\mathrm{cm}^{-1}$.}

\label{radiation_matter_darkenergy}
\end{figure*}
\subsection{The case when the CDM sector, baryonic matter, radiation, and dark energy are present}

We now analyze the evolution of a multi-component universe consisting of radiation (R), cold dark matter (M), baryonic matter (B) and dark energy (DE), beginning from an early radiation-dominated epoch and continuing up to the present day, characterized by a scale factor $a_{0}=1$.  The cold dark matter is modeled by the oscillating scalar field $\phi$ while the baryonic matter is treated as pressureless dust with its energy density scaling as $1/a^{3}$. To ensure radiation domination at early times, we set initial conditions at a scale factor $a_{i}=2.24 \times 10^{-4}$, corresponding to a time well before the epoch of radiation–matter equality. Here, the scale factor will be dimensionless because the universe is flat. At this initial time, $t=t_{i}$ the radiation energy density is $\rho_{\text{R}_{i}}=\rho_{\text{R}_{0}}/a_{i}^{4}=5.56\times10^{10}~u^{2}$, the baryonic matter density is $\rho_{\text{B}_{i}}=\rho_{\text{B}_{0}}/a_{i}^{3}=7.02\times10^{9}~u^{2}$ , and the dark matter energy density is $\rho_{\text{M}_{i}}=\rho_{\text{M}_{0}}/a_{i}^{3}=3.74\times10^{10}~u^{2}$. Since dark matter is modeled using the oscillating scalar field $\phi$, we determine its initial conditions to ensure that its averaged out energy density $\langle \rho \rangle$  given by Eq.~(\ref{rhsfried14n}) matches $\rho_{\text{M}_{i}}$ at $t=t_{i}$. Specifically, we choose: $\phi_{+}(t_{i}) = \phi_{-}(t_{i}) = 10^{-6},$ $\dot{\phi}_{+}(t_{i}) = \dot{\phi}_{-}(t_{i}) = -\frac{3H_i}{2} \times 10^{-6}~u$, in accordance with Eq.~(\ref{initcn}). Here, $H_i = \left. \frac{\dot{a}}{a} \right|_{t=t_{i}}=\frac{\dot{a_{i}}}{a_{i}}=1.82\times10^{5}~u$ denotes the Hubble parameter at the initial time. Despite the large initial value of the Hubble parameter, the condition $m \gg H$ remains satisfied throughout the evolution considered here, thus validating the applicability of our method.

For dark energy, we have considered two distinct models:
\begin{enumerate}
    \item Cosmological Constant ($\Lambda$): Representing a constant vacuum energy density $\rho_{\Lambda}$.
    \item Dynamical Scalar Field ($\psi$): This scalar field can represent either quintessence
($\epsilon=+1$) or phantom energy ($\epsilon=-1$) as stated before.
\end{enumerate}

We begin by demonstrating the validity of our method in a multicomponent universe consisting of the oscillating ultralight scalar field $\phi$ (representing dark matter), baryonic matter (B), radiation (R), and a cosmological constant ($\Lambda$). The Friedmann equation employed for this scenario is:
\begin{eqnarray}
  \left(\frac{\dot{a}}{a}\right)^2 = \frac{1}{3}\left[\langle \rho \rangle + \frac{\rho_{R_{0}}}{a^{4}}+\frac{\rho_{B_{0}}}{a^{3}}+\rho_{\Lambda}\right]\,,
\label{minimal1}
\end{eqnarray}
which is solved in conjunction with Eqs.~(\ref{gen1}) and (\ref{gen2}) from initial time $t_{i}$ to present epoch $t_{0}$.

The resulting cosmological evolution is shown in Fig.~[\ref{radiation_matter_darkenergy}]. Numerical analysis confirms that the coherently oscillating ultralight scalar field closely replicates the dynamical behavior of the cold dark matter model, with the averaged out scalar field EoS parameter, $\omega$, remaining effectively zero throughout. In Fig.~[\ref{radiation_matter_darkenergy1}] we plot the energy density parameters of the individual components as,  baryonic matter (B), radiation (R), scalar field dark matter ($\phi$) and the cosmological constant ($\Lambda$) as a function of $\log_{10} (1+\text{z})$, where the redshift z, plotted in decreasing order to represent the forward progression of cosmic time. This visualization effectively captures the time evolution of the density parameters for each component.

The evolution of the multicomponent universe follows the standard cosmological sequence. It begins in a radiation-dominated phase. As expansion proceeds, the matter energy density surpasses the radiation density, marking the onset of the matter-dominated era. Eventually, dark energy, represented by the cosmological constant, dominates and drives late-time accelerated expansion. Importantly, the oscillating ultralight scalar field used to represent dark matter replicates the standard behavior.  It reproduces the correct redshift values for both the radiation–matter equality and the matter–dark energy equality, consistent with what is expected in the case of cold dark matter. The effective EoS in Fig.~[\ref{radiation_matter_darkenergy6}] evolves accordingly, starting from a positive value during the radiation era and gradually approaching $-1$, as the system transitions into the dark energy-dominated phase.


Instead of employing the cosmological constant, one can also work with the dynamical dark energy model where a scalar field $\psi$ models the dark energy sector. This scalar field can be either the quintessence or the phantom field. The initial conditions for dark matter (modeled by an oscillating ultralight scalar field $\phi$), baryonic matter (B) and radiation (R) remain unchanged from the previous case and the initial conditions and parameters defining the dark energy density are set to the same values as discussed in the concluding subsection of the previous section. In this scenario, the Friedmann equation takes the form: 
\begin{eqnarray}
  \left(\frac{\dot{a}}{a}\right)^2 = \frac{1}{3}\left[\langle \rho \rangle + \frac{\rho_{(R)_{0}}}{a^{4}}+\frac{\rho_{(B)_{0}}}{a^{3}}+\rho_{(\psi)}\right]\,.
\label{minimal3}
\end{eqnarray}
The complete cosmological dynamics of this system from initial time $t_{i}$ to present epoch $t_{0}$ is then obtained from the full system of equations which include Eqs.~(\ref{gen1}), (\ref{gen2}), (\ref{kGeqforpsi}), and (\ref{minimal3}).

The Fig.~[\ref{radiation_matter_darkenergy}] illustrates the results when the dark energy is modeled by quintessence and phantom fields. The numerical results in Figs.~[\ref{radiation_matter_darkenergy2}, \ref{radiation_matter_darkenergy3}, \ref{radiation_matter_darkenergy4}] show a broadly similar behavior to the case of the cosmological constant case as expected; some subtle differences appear in the evolution of the effective equation of state (EoS) parameter,  $\omega_{\text{eff}}$, in Fig.~[\ref{radiation_matter_darkenergy6}], which now deviates from the value it had in the cosmological constant model and evolves differently depending on the chosen scalar field model. Also, the dark energy EoS parameter in Fig.~[\ref{radiation_matter_darkenergy5}] is no longer strictly constant, as it is for a cosmological constant, but evolves with time, taking values greater than $-1$ for quintessence and less than $-1$ for phantom fields.

From the discussion presented above, it is seen that, the oscillating ultralight scalar field dark matter consistently reproduces the evolution expected from CDM throughout this evolution, which makes our generalization scheme valid in a multicomponent universe. This reinforces the reliability of using coherently oscillating ultralight scalar fields as dark matter analogues, even when the dark matter is not the dominant component.
\subsection{The case when CDM, radiation, and dark energy are present and the CDM sector decays into the radiation sector}

In the case, where the scalar field decays into radiation, we begin with the same initial values as in the previous subsection, corresponding to a radiation domination epoch. A decay constant of $\Gamma=10^{-4}~u$ is introduced to govern energy transfer from the scalar field to radiation. In this scenario, the dark matter component, represented by the oscillating ultralight scalar field, gradually decays into radiation in the presence of baryonic matter and dark energy. We first analyze the evolution of a system where dark energy is modeled by the cosmological constant, and then proceed to the case where dark energy is described by a dynamical scalar field.

The time evolution for the first case is governed by Eqs.~(\ref{radend}), (\ref{gen1n}), (\ref{gen2n}), along with the modified Friedmann equation: 
\begin{eqnarray}
  \left(\frac{\dot{a}}{a}\right)^2 &=& \frac{1}{3}\left[\langle \rho \rangle +\rho_{R}+\frac{\rho_{B_{0}}}{a^{3}}+\rho_{\Lambda}\right]\,.
\label{nonminimal1}
\end{eqnarray}
The resulting time evolution of the various cosmological variables from this coupled system of equations can be checked similarly to the previous case. Here, the oscillating ultralight scalar field dark matter decays into radiation under the influence of a cosmological constant. Despite the decay, the effective EoS parameter of the dark matter component remains that of pressureless dust. Importantly, it can be verified and is also shown in the appendix (\ref{cosmological_parameter_appa}) in Fig.~(\ref{nonminimal_radDE}) that the density parameters of the individual components evolve such that their present-day values are consistent with observations, indicating that even a small coupling between dark matter and radiation can yield a viable cosmological evolution. This shows that standard cosmological dynamics can admit a small value of $\Gamma \sim 10^{-4}u$ which is equivalent to  $10^{-37}$\,eV in natural units for comparison. This small decay rate from the CDM particles to photons does not affect the results of cosmological dynamics till now. We will come back to this topic at the end of this subsection. Next, we present the results assuming a dynamic dark energy sector.

Here, we separately employ quintessence and phantom-like scalar fields to model dark energy. The governing equations for the system are given by Eqs.~(\ref{radend}), (\ref{gen1n}), (\ref{gen2n}) with (\ref{kGeqforpsi}) along with the modified Friedmann equation,
\begin{eqnarray}
  \left(\frac{\dot{a}}{a}\right)^2 &=& \frac{1}{3}\left[\langle \rho \rangle +\rho_{R}+\frac{\rho_{B_{0}}}{a^{3}}+\rho_{\psi}\right]\,.
\label{nonminimal2}
\end{eqnarray}
Again, one can examine the time evolution of the various cosmological variables for the system described, and the behavior exhibited closely parallels that observed in the scenario where cosmological constant models are used to describe dark energy. As in the minimally coupled case, both the dark energy equation of state (EoS) driven by the scalar field and the overall effective EoS of the system depend on the type of scalar field used to model the dynamic dark energy sector. This can be easily verified and has been shown in the appendix (\ref{cosmological_parameter_appa}). These results show that the generalization of the coherent oscillation model of the ultralight scalar field model works properly in the cosmological sector. In this paper, we have not compared our results with modern cosmological data; in a future publication, we would like to do that and set a more accurate phenomenological bound on $\Lambda$. In the present calculations, we have used a value of $\Gamma$ which is sufficiently smaller in value when compared to the Hubble parameter value. Technically, we are in the weak dissipative regime where the scalar field energy weakly transforms into radiation. In the far future, the value of the Hubble parameter will reduce and then $\Gamma\sim H$, and then the above results may change. If such a regime arises at all, then we can have the conversion of the whole cold dark matter energy into radiation. To ensure numerical consistency, we have also verified the energy balance for all decay runs by evaluating the residual of the radiation continuity equation, $\mathcal{R}=\dot{\rho}_{R} + 4H\rho_{R} - \Gamma\langle\dot{\phi}^2\rangle.$ The residual remains negligible throughout the evolution, confirming that the total energy transfer between the scalar field and radiation is accurately conserved. Further details are provided in Appendix~\ref{Radiation_continuity_appa}, along with the corresponding results shown in Fig.~\ref{radiation_continuity}.

\begin{figure*}
 \subfigure[ \scriptsize Evolution of the normalized scale factor ($a/a_{\text{v}}$) with normalized time ($t/t_\text{v}$) for different dark energy models.]
{\includegraphics[width=68mm,height=33mm]{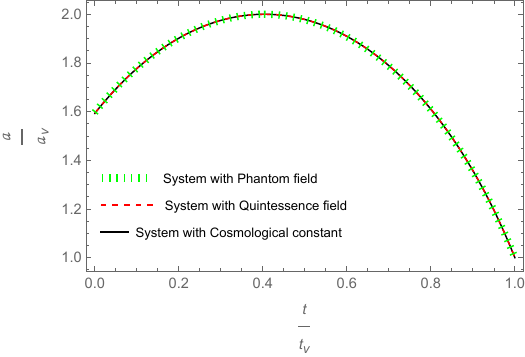}\label{nonminimal_radDECu1}}
\hspace{0.1cm}
\subfigure[\scriptsize Evolution of dark energy equation of state parameter ($\omega_{\text{DE}}$) with normalized time ($t/t_\text{v}$) for different dark energy models.]
{\includegraphics[width=68mm,height=33mm]{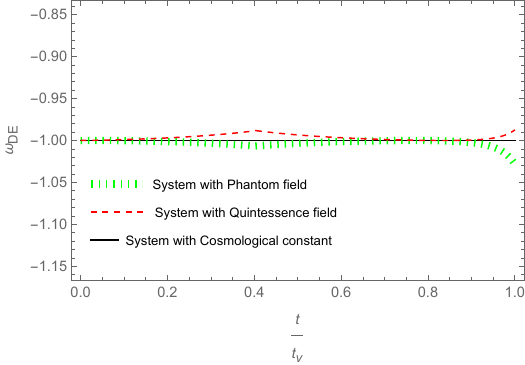}\label{nonminimal_radDECu2}}
\hspace{0.1cm}
\subfigure[\scriptsize Evolution of various normalized energy densities ($\rho/\rho_{\text{v}}$) with normalized time ($t/t_{\text{v}}$) in a cosmological model with dark energy modeled as the cosmological constant ($\Lambda$).]
{\includegraphics[width=68mm,height=33mm]{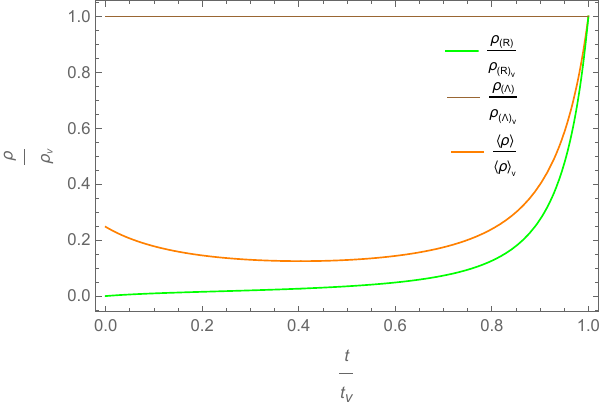}\label{nonminimal_radDECu3}}
\hspace{0.1cm}
\subfigure[\scriptsize Evolution of various normalized energy densities with normalized time ($t/t_{\text{v}}$) in a cosmological model with dark energy modeled as a quintessence scalar field ($\psi$).]
{\includegraphics[width=68mm,height=33mm]{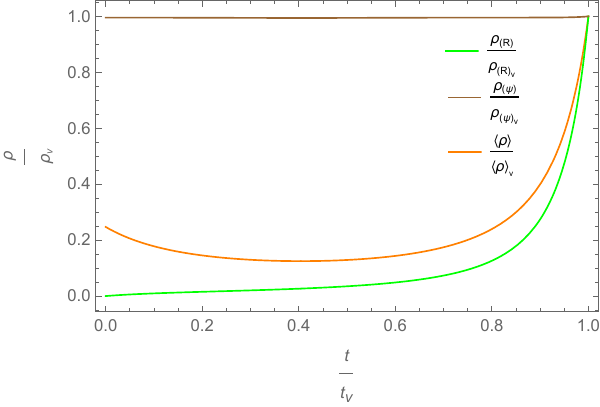}\label{nonminimal_radDECu4}}
\hspace{0.1cm}
\subfigure[\scriptsize Evolution of various normalized energy densities with normalized time ($t/t_{\text{v}}$) in a cosmological model with dark energy modeled as a phantom scalar field ($\psi$).]
{\includegraphics[width=68mm,height=33mm]{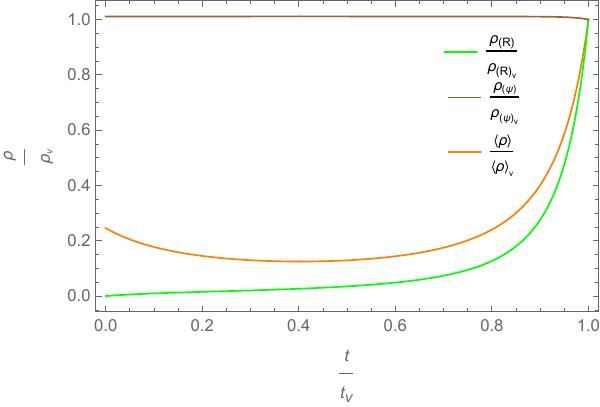}\label{nonminimal_radDECu5}}
\hspace{0.1cm}
\subfigure[\scriptsize Evolution of the effective equation of state parameter $\omega_{\text{eff}} = (\langle \rho \rangle + \tilde{\rho}) / (\langle P \rangle + \tilde{P})$ of the system with normalized time ($t/t_\text{v}$) for different dark energy models. Dark matter quantities are time-averaged ($\langle \cdot \rangle$), while dark energy quantities are not.]
{\includegraphics[width=68mm,height=33mm]{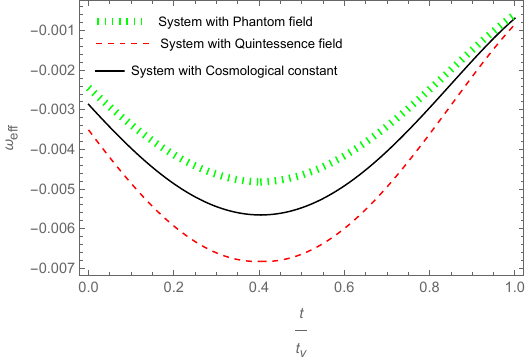}\label{nonminimal_radDECu6}}
\hspace{0.1cm}
\subfigure[\scriptsize Evolution of time-averaged dark matter equation of state parameter ($\omega$) with normalized time ($t/t_\text{v}$) for different dark energy models.]
{\includegraphics[width=68mm,height=33mm]{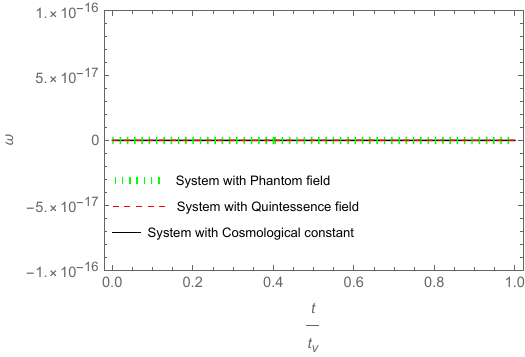}\label{nonminimal_radDECu7}}
\hspace{0.1cm}
\renewcommand{\figurename}{\footnotesize Figure}
\caption{\footnotesize Time evolution of system variables for an overdense region treated as a closed system with positive curvature ($k = 1$). Radiation $(R)$ is sourced by the decay of an oscillating ultralight scalar field ($\phi$), representing cold dark matter, with mean energy density $\langle \rho \rangle$, and a decay rate $\Gamma = 10^{-4}~u$. Dark energy is incorporated through three distinct models: (1) a cosmological constant with $\Lambda = 1.08~u^2$; (2) a quintessence-like scalar field ($\psi$) with potential $V(\psi) = V_{m} e^{-\lambda \psi}$, where $V_{m} = 1.38~u^2$; and (3) a phantom-like scalar field ($\psi$) with the same potential form but $V_{m} = 0.91~u^2$. For both scalar field models, the parameter $\lambda = 1$ and initial conditions $\psi(t_i) = 0.0001$, $\dot{\psi}(t_i) = 0.0001~u$. All variables are evaluated up to the virialization time $t_\text{v}$. Dark matter $\phi$ has mass $m = 1.93\times 10^{11}\, u$ with initial conditions $\phi_\pm(t_i) = 10^{-6}$, $\dot{\phi}_\pm(t_i) = -\frac{3H_i}{2} \times 10^{-6}~u$, where $a_i = 2.24\times10^{-4}$ and $H_i = 1.82\times10^5~u$. The initial radiation energy density $\rho_{(R)_{i}} = 0.00001~u^{2}$. All quantities are expressed in geometrized units with \(u = 10^{-28}~\mathrm{cm}^{-1}\).}
\label{nonminimal_radDECu}
\end{figure*}
\section{Application of our method in gravitational collapse}
\label{gravcol}

In this section, we demonstrate that the generalization of the coherent oscillation of the ultralight scalar field model can behave similarly to the cold dark matter sector, even in the context of gravitational collapse. Such a collapsing scenario was recently studied in Ref.~\cite{Saha:2024irh} where the authors used an approximated version of the coherent oscillation of the ultralight scalar field model. Unlike the previous work, in the present article, we will use the exact generalization of the model with the proper initial conditions in the interior patch of the spacetime. Before we proceed, it is appropriate to state here that the gravitational collapse model studied in this paper serves an illustrative purpose where we particularly concentrate on the scalar field and CDM sector equivalence and the specific role of the initial condition as given in Eq.~(\ref{initcn}). In this paper, we do not work out explicitly how the collapsing spacetime is matched with an external static, spherically symmetric spacetim,e as we do not expect any new features in the matching sector when we compare or work with the previous works involving gravitational collapse in Refs.~\cite{Saha:2024irh, Saha:2024xbg, Saha:2023zos}. The effect of the generalization of the equivalence between the CDM sector and the oscillating scalar field sector primarily affects the internal collapsing spacetime region. We will specify more about the matching conditions at the end of this section.

Here we investigate the dynamics of gravitational collapse in a system where dark matter decays into radiation in the presence of dark energy. The evolution is studied in a closed FLRW spacetime with positive spatial curvature $(k=1)$. Dark matter is modeled by the coherently oscillating ultralight scalar field. We focus on the behavior of a small, spherically symmetric overdense region embedded in the expanding universe. Due to its initially higher density relative to the background, this region gradually decouples from the Hubble flow, experiences a slowdown in expansion, and eventually reaches a turnaround point, initiating its gravitational collapse. Our analysis, however, does not proceed to the final singularity of the collapse. Instead, it concentrates on the stage when the system attains virial equilibrium and stabilizes. The primary motivation for introducing a phenomenological mechanism to halt the collapse is rooted in the observational evidence of large-scale structures in the universe, which are believed to form via such virialization processes. A collapsing region must, therefore, reach a virialized state at some stage in its evolution.

In our model, we employ the conventions used in the standard \textit{top-hat collapse} framework~\cite{Gunn:1972sv}, wherein the overdense region initially expands along with the background universe but at a slower rate. Eventually, it reaches a maximum expansion radius—known as the turnaround radius, $R_{\text{max}}$. At this point, the kinetic energy of the overdense region momentarily vanishes, and the total energy is purely gravitational. The total gravitational potential energy at turnaround is given by: $ E_T = V_T = -\frac{3M^2}{5R_{\text{max}}},$ where $M$ is the total mass within the overdense region \cite{Shapiro:1998zp}. As the collapse continues, the region eventually virializes. At virial equilibrium, the virial theorem implies that the total kinetic energy, $E_K$, and the potential energy, $V_T$, satisfy: $ E_K = -\frac{1}{2}V_T,$ which yields the total energy at virialization as: $ E_T = E_K + V_T = \frac{V_T}{2}.$  Since the total energy is conserved during the evolution, equating the expressions for $E_T$ at turnaround and virialization gives:
$$\frac{R_{\text{v}}}{R_{\text{max}}} = \frac{1}{2},$$
which we adopt to define the virialized state of the collapsing region. In our setup, we define the scale factor of the overdense region at the time of virialization to be half of its value at turnaround, i.e. $\frac{a_{\text{v}}}{a_{\text{max}}} = \frac{1}{2}$, where the subscript "v" denotes the virialized state.
Although we define the virialization of our system as defined above, our system is more complicated than the traditional top-hat collapse model because we have a multicomponent overdense region, and out of all the various components, only the CDM-like sector collapses. Here we assume that the collapse process comes to a halt when the CDM sector virializes and $a_{\rm v} \sim a_{\rm max}/2$, the other components of the overdense patch redistribute their energy densities during this process of virialization. This assumption was used previously to define the virialization process of a multicomponent overdense patch, where the CDM-like sector primarily collapses, in Ref.~\cite{Saha:2024irh}.

In the present case, we assume that the overdense region initially does not contain baryonic matter or radiation; radiation is generated dynamically through the decay of dark matter. Thus, we set the initial radiation density at $t=t_{i}=0$ to zero, $\rho_{R_{i}} = 0$, in principle. For numerical stability, we begin with a small seed value, $\rho_{R_{i}} = 0.00001~u^{2}$. As before, dark energy is modeled in two distinct ways: as a cosmological constant and as a scalar field $\psi$, in both cases, we use the initial conditions and parameter values as given in subsection \ref{incs}. This general framework allows us to study any epoch by choosing different values of the scale factor. We select $a_{i}=0.1~u^{-1}$, here scale factor $a$ is not dimensionless like the before case of the flat universe with $k=0$, it has dimensions of length because we are working with $k=1$. The initial conditions for the oscillating scalar field at, $t_{i} = 0$  are given as: $\phi_{+}(t_{i}) = \phi_{-}(t_{i}) = 10^{-10},$ $\dot{\phi}_{+}(t_{i}) = \dot{\phi}_{-}(t_{i}) = -\frac{3H_i}{2} \times 10^{-10}~u$, with $H_i=0.454~u$. All quantities are rendered dimensionless by normalizing with respect to their values at virialization: $a/a_{v}$, $t/t_{v}$, and $\rho/\rho_{v}$. The mass of the ultralight scalar field ($m$) and the coupling constant ($\Gamma$) have the same value they had in the previous sections.

In the case where dark energy is represented by a cosmological constant, the Friedmann equation takes the form:
\begin{eqnarray}
  \left(\frac{\dot{a}}{a}\right)^2 &=& \frac{1}{3}\left[\langle \rho \rangle + \rho_{\text{R}}+\rho_{\Lambda}\right]-\frac{1}{a^{2}}\,.
\label{nonminimal3}
\end{eqnarray}
The evolution of the system is governed by  Eqs.~(\ref{radend}), (\ref{gen1n}), (\ref{gen2n}), in conjunction with the above Friedmann equation. The resulting dynamics are illustrated in Fig.~[\ref{nonminimal_radDECu}]. In the figure, we have rendered all the dimensional quantities dimensionless by scaling their values with respect to the values these quantities will have at virialization. The plots show that the generalization of the coherent oscillation of the ultralight scalar field model excellently represents the CDM-like sector, which predominantly collapses. During this collapse, some scalar field energy is transformed into radiation. The matter EoS shows that the multicomponent overdense region still behaves like dust. 

For the case where dark energy is modeled by a scalar field, we solve the system using Eqs.~(\ref{radend}), (\ref{gen1n}), (\ref{gen2n}) with (\ref{kGeqforpsi}) along with the modified Friedmann equation:
\begin{eqnarray}
  \left(\frac{\dot{a}}{a}\right)^2 &=& \frac{1}{3}\left[\langle \rho \rangle + \rho_{\text{R}}+\rho_{\psi}\right]-\frac{1}{a^{2}}\,.
\label{nonminimal4}
\end{eqnarray}
Fig.~[\ref{nonminimal_radDECu}] shows both the evolution of the system when dark energy is modeled as quintessence and phantom-like scalar fields. In the plots, we see that the coherently oscillating ultralight scalar field sector works properly, whereas the EoS for the quintessence and phantom fields are slightly different, as expected. Like the previous case, we have rescaled all dimensional quantities by the values they will have at virialization. The results show that some amount of radiation will be produced in the overdense patches. 

In all of the above cases, the overdense patch does not act like a closed universe, as there are components that do not collapse, and consequently, there will be nonzero pressure on the boundary of the overdense patch. To tackle this problem, one has to match a different spacetime on the boundary of the overdense patch, which can take care of the imbalanced pressure at the boundary. One in general uses the generalized Vaidya spacetime as the outer spacetime, as explicitly shown in the  Refs.~\cite{Saha:2024irh, Saha:2024xbg, Saha:2023zos}. In this paper, we do not reproduce those results as the main purpose of this paper is not the junction conditions but the validity of the coherently oscillating ultralight scalar field model. Before we end, we must point out that gravitational collapse in a multicomponent world, where dark energy is one of the components, is sensitive to initial conditions. In general, one does not get collapsing solutions for any arbitrary initial condition. This is a separate topic, and we will not discuss it further here. Interested readers who want to know more about the initial conditions of gravitational collapse may consult the last set of references cited above in this paragraph. 

\section{Results and discussions}

In the present paper, we have proposed an approximate equivalence between the coherently oscillating ultralight scalar field model and the standard CDM model in the case when there are multiple components in the cosmological system. The equivalence becomes nearly exact in the limit $m \gg H$ where $m$ designates the mass of the ultralight scalar field and $H$ is the Hubble parameter. The equivalence requires a specific form of initial conditions on the ultralight scalar field sector. Only when these initial conditions are satisfied then the equivalence work perfectly. Previous authors who have used the equivalence did not use the specific initial conditions given in this work, and consequently, their works are only approximations of the real result.   

The initial conditions on the oscillating field required for the aforementioned equivalence, as given in Eq.~(\ref{initcn}), turn out to be a constraint which has dynamical significance. The initial relationships satisfied by the various components of the ultralight scalar field at $t=0$, in Eq.~(\ref{initcn}), hold approximately (with a precision depending on how large $m/H$) even at any finite time.  Any arbitrary initial condition on the oscillating scalar field will not reproduce the CDM-like features expected from the equivalence. This observation has interesting consequences. Before discussing the consequences of this observation, we have to specify the limits of the equivalence.

In our calculation, we have worked under the assumption that $m \gg H$, which ensures that the scalar field undergoes rapid oscillations and can be effectively treated as cold dark matter. This assumption does not hold universally, especially when we consider earlier epochs in the Universe's history corresponding to larger redshift values $z$. As we go further back in time, the Hubble parameter $H$ increases and eventually becomes comparable to the scalar field mass $m=3.8\times 10^{-22}$\,eV. Specifically 
at $z \sim  5.3144 \times 10^6$, we find that $H = m$, marking the boundary beyond which our assumption $m \gg H$ breaks down. Therefore, our calculation remains valid only for redshift $z \lesssim 5.31 \times 10^6$, and the scalar field would not exhibit dark matter-like behavior before or near this redshift. The approximate value of $ z\sim 10^6$ above which the equivalence fails matches with the corresponding value of $z$ reported in Ref.~\cite{Mishra:2017ehw}. The above discussion shows that near about $z \sim 10^5$ the equivalence discussed in this paper becomes ineffective because the two different timescales $1/m$ and $1/H$ come close to each other. For redshifts above $10^5$ one cannot assume that the scalar field oscillates during the cosmological time scale, and consequently, one has to solve the scalar field equation conventionally without using any averaging procedure. As a result of this discussion, we see that the initial condition discussed in the previous paragraph becomes interesting. One can solve the ultralight scalar field equation in the deep radiation dominated phase ($z>10^5$) using some other methods, but if one demands that the same scalar field must work as the CDM candidate then the ultralight scalar field has to satisfy the restrictive conditions in Eq.~(\ref{initcn})  at some later time, $t_i$, when $m\gg H$. In that case, the equivalence starts to be operational for $t>t_i$. The question remains, which of the initial conditions in the pre-oscillating phase leads to the specific kind of conditions one has in Eq.~(\ref{initcn}) in the future? We would like to address this issue in the near future.

We have seen that the equivalence of the coherent oscillating ultralight scalar field model and the CDM sector allows the ultralight scalar field to have a minute decay rate $\Gamma = 10^{-4}\,u$, expressed in natural units as $10^{-37}\,\mathrm{eV}$. Although this decay rate does not produce a significant difference from the standard cosmological development, this small value of $\Gamma$ can have a nontrivial effect in gravitational collapse. The dark matter collapse locally will produce faint radiation. Moreover, in the future $H \sim \Gamma$ (signaling the onset of a strong dissipative regime) and consequently all the CDM will start to convert into radiation. If we allow the small $\Gamma$ to exist, then we can predict that the universe will convert all its CDM component into radiation in the future. 

With the initial conditions or constraint on the ultralight scalar field we have shown that the equivalence between the oscillating ultralight scalar field model and the CDM sector, for the multicomponent universe,  works reasonably well and can reproduce the proper cosmological phases with proper predictions of the redshifts of transition from one phase to the other phase. We have also shown how the equivalence can be used in the case of gravitational collapse in the late phase of the universe. These discussions show the effectiveness of our proposed equivalence. One can now look at the fate of cosmological perturbations using the aforementioned equivalence.

\appendix

\section{The details of the averaging process}
\label{appa}

In our case $\sin \psi$, $\cos \psi$, $\sin^2 \psi$, $\cos^2 \psi$ are all oscillating functions with period $2\pi/m$. 
Using the results:
$\langle \sin^2 \psi \rangle = \langle \cos^2 \psi \rangle = 1/2$, and 
$\langle \sin \psi \cos\psi \rangle =0$  we have:
\begin{eqnarray}
\langle\dot{\phi}^2\rangle=&& \frac12 \left[\dot{\phi}_+^2 + \dot{\phi}_-^2 + m^2({\phi}_+^2 + {\phi}_-^2) \right]\nonumber\\&& + m(\dot{\phi}_-\phi_+ - \dot{\phi}_+ \phi_-)\,,
\label{phi0d2}
\end{eqnarray}
and
\begin{eqnarray}
m^2\langle \phi^2 \rangle =\frac12(\phi_+^2 + \phi_-^2)m^2\,.
\label{phi02}
\end{eqnarray}
As a consequence, we have
\begin{eqnarray}
\langle\dot{\phi}^2 + m^2\phi^2\rangle =&&\frac12 (\dot{\phi}_+^2 + \dot{\phi}_-^2) + m^2({\phi}_+^2 + {\phi}_-^2) \nonumber\\&&+ m(\dot{\phi}_-\phi_+ - \dot{\phi}_+ \phi_-)\,. 
\label{rhsfried1}
\end{eqnarray}
From Eq.~(\ref{phitime}) we get the condition:
\begin{eqnarray}
(\dot{\phi}_-\phi_+ - \dot{\phi}_+ \phi_-)=0\,,
\label{initcons}
\end{eqnarray}
which implies
\begin{eqnarray}
\langle\dot{\phi}^2 + m^2\phi^2\rangle =\frac12 (\dot{\phi}_+^2 + \dot{\phi}_-^2) + m^2({\phi}_+^2 + {\phi}_-^2)\,. 
\label{rhsfried12}
\end{eqnarray}
Using Eq.~(\ref{mordered3}) and Eq.~(\ref{phitime}) we can write
\begin{eqnarray}
\langle\dot{\phi}^2 + m^2\phi^2\rangle&& =\frac98 (C_+^2 + C_-^2)\frac{H^2}{a^3} + m^2(C_+^2 + C_-^2)\frac{1}{a^3}\nonumber\\&&=m^2(C_+^2 + C_-^2)\frac{1}{a^3}\left(1+ \frac98 \frac{H^2}{m^2}\right)\,. 
\label{rhsfried13}
\end{eqnarray}
The above results allow us to write the energy density of the system as:
\begin{eqnarray}
\langle\rho\rangle=\frac12\langle\dot{\phi}^2 + m^2\phi^2\rangle =&&\frac12 m^2(C_+^2 + C_-^2)\frac{1}{a^3}\nonumber\\&&\left(1+ \frac98 \frac{H^2}{m^2}\right)\,. 
\label{arhsfried14}
\end{eqnarray}
The pressure of the system is given by (From Eq.~(\ref{phi0d2}) and Eq.~(\ref{phi02}))
\begin{eqnarray}
  \langle P \rangle &&= \frac12 \langle\dot{\phi}^2 - m^2\phi^2\rangle = \frac14  (\dot{\phi}_+^2 + \dot{\phi}_-^2)\nonumber\\&& = \frac{9}{16}m^2 (C_+^2 + C_-^2)\frac{1}{a^3}\left(\frac{H^2}{m^2}\right)\,.
\label{apressb}  
\end{eqnarray}
The ratio of the pressure and the energy density of the system then yields:
\begin{eqnarray}
\frac{\langle P \rangle}{\langle\rho\rangle}= \frac{\frac98\frac{H^2}{m^2}}{\left(1+ \frac98 \frac{H^2}{m^2}\right)}\,.
\label{abeos}
\end{eqnarray}
In the limit $m \gg H$, we see that the above ratio tends to zero, specifying an effective equation of state (EoS) of dust. 

Using Eq.~(\ref{rhsfried14}) the first Friedmann equation gives
$$\left(\frac{\dot{a}}{a}\right)^2 = \frac{1}{3}\langle \rho \rangle = \frac{1}{6}m^2(C_+^2 + C_-^2)\frac{1}{a^3}\left(1+ \frac98 \frac{H^2}{m^2}\right)\,.$$
Because $m \gg H$, we can neglect the term $H^2/m^2$ and write the solution of the above equation as:
\begin{eqnarray}
a(t)=a_{i}\left[1 + A(t-t_i)\right]^{2/3}\,,
\label{aasoln}
\end{eqnarray}
where $a_{i}$ is the scale-factor at $t=t_{i}$ and
\begin{eqnarray}
A^2 \equiv \frac{3m^2}{8a_i^3}(C_+^2 + C_-^2)\,.
\label{aAdef}
\end{eqnarray}
This is the matter-dominated universe solution. The second Friedmann equation can be written as
\begin{eqnarray}
\frac{\ddot{a}}{a}=-\frac{1}{6}\langle \rho + 3P \rangle\,.
\label{secfried}
\end{eqnarray}
We have seen previously that when $m \gg H$ we have $\langle P \rangle \ll \langle \rho \rangle$.
In this limit, we can safely neglect the pressure term in the above equation. In this limit the above equation is satisfied for the scale-factor expression given in Eq.~(\ref{asoln}).

\section{Formalism and some more details about the new solutions}
\label{formal}

If we consider a single homogeneous scalar field that seeds the spacetime and which is oscillating much faster compared to the cosmological time scale, then the Cauchy data would be: $(\Sigma,~h_{ab},~ \mathcal{K}_{ab},~\phi(0),~\dot{\phi}(0))$, where $h_{ab}, \mathcal{K}_{ab}$ are the induced metric and the second fundamental form on $\Sigma$, respectively. However, one has to choose the data in such a way that it must satisfy the following constraint equations on $\Sigma$:
\begin{eqnarray}
\label{hamConst}
^{(3)}R+(tr(\mathcal{K}_{ab}))^2-\mathcal{K}^2=2\langle \rho \rangle\, ,\\
\tilde{\nabla}_a(\mathcal{K}^{ab}-h^{ab}~tr(\mathcal{K}))=\langle j^b \rangle,
\label{momeConst}
\end{eqnarray}
where $^{(3)}R$ is Ricci scalar of co-dimension one spacelike $\Sigma$, $\tilde{\nabla}_a$ is the covariant derivative on that surface, $\mathcal{K}^2=\mathcal{K}_{ab}\mathcal{K}^{ab}$, $\langle\rho \rangle=-T^0_0=-G^0_0$, and the induced momentum density $\langle j_a \rangle = T_{\alpha\beta}~n^\alpha~ e^\beta_a$ where $n^\alpha$ and $e^\beta_a$ are the timelike normal and spacelike tangent to the surface $\Sigma$, respectively.  Here, the angular brackets specify the time average over the time period of scalar field oscillations. It should be noted that in this paper, the roman indices $\{a,b,c,...\}$ are used to represent the spatial indices while the Greek indices $\{\alpha,\beta,\gamma,...\}$ denote the spacetime indices. Now, in our case, the induced metric on the Cauchy surface is:
\begin{eqnarray}
h_{ab} = \begin{pmatrix}
\frac{a^2(0)}{1-kr^2} & 0 & 0 \\
0 & r^2a^2(0) & 0\\
0 & 0 & r^2\sin^2\theta ~a^2(0)
\end{pmatrix},
\end{eqnarray}
where the $t=0$ characterizes the spacelike property of the Cauchy surface $\Sigma$. Now, we can compute ($\mathcal{K}_{ab}$) and $^{(3)}R$ on $\Sigma$. Using these and also using the first constraint Eq.~(\ref{hamConst}), we get the first Friedmann Eq.~(\ref{cfried1}). It can be verified that the second constraint, Eq.(\ref{momeConst}), is trivially satisfied for the aforementioned induced metric on $\Sigma$. Therefore, here, the Cauchy problem consists of solving the evolution equation of the metric, i.e., the second Friedmann equation (Eq.~(\ref{cfried2})) and the evolution equation for $\phi(t)$ (Eq.~(\ref{canscbt})) or equivalently the evolution equations for $\phi_+(t)$ and $\phi_-(t)$ (Eq.~(\ref{gen1}) and Eq.~(\ref{gen2}) respectively) with the Cauchy data $(\Sigma,~h_{ab},~ \mathcal{K}_{ab},~\phi(0),~\dot{\phi}(0))$ or equivalently $(\Sigma,~h_{ab},~ \mathcal{K}_{ab},~\phi_{\pm}(0),~\dot{\phi}_{\pm}(0))$  that satisfy the two constraint equations mentioned above. It can be shown that the propagation of the constraint equations along time $(t)$ is evident with this construction. 

Now, apart from the constraint equation Eq.~(\ref{hamConst}), there is no intrinsic constraint on scalar fields ($\phi_{\pm}(0)$) and their derivatives $(\dot{\phi}_{\pm}(0))$ on $\Sigma$. Therefore, we are free to choose five initial values out of six  $(a(0),~ H(0),~\phi_{\pm}(0),~\dot{\phi}_{\pm}(0))$, where $a|_{\Sigma},~ H|_{\Sigma}$ are related to $h_{ab},~ \mathcal{K}_{ab}$, respectively. For a specific model of scalar fields or space-time dynamics, additional constraint equations may arise that must be satisfied on the Cauchy surface. Next, we discuss the requirement of imposing additional constraints on the scalar fields and their derivatives on $\Sigma$ to ensure that the scalar field closely mimics the behavior of cold dark matter (CDM). 

\section{Additional Diagrams}\label{Diagrams}

\subsection{Evolution of Cosmological Parameters when CDM decays into radiation}\label{cosmological_parameter_appa}

\begin{figure*}
\subfigure[Evolution of density parameters ($\Omega$) with redshift (z) in a cosmological model with dark energy modeled as the cosmological constant ($\Lambda$).]
{\includegraphics[width=85mm,height=50mm]{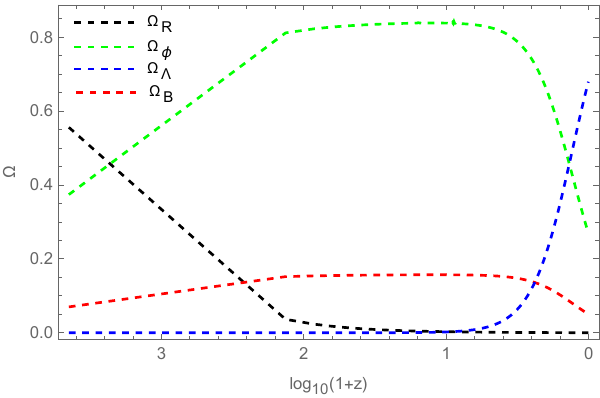}\label{nonminimal_radDE1}}
\hspace{0.2cm}
\subfigure[Evolution of the density parameters ($\Omega$) with redshift (z) in a cosmological model with dark energy modeled as a quintessence scalar field ($\psi$).]
{\includegraphics[width=85mm,height=50mm]{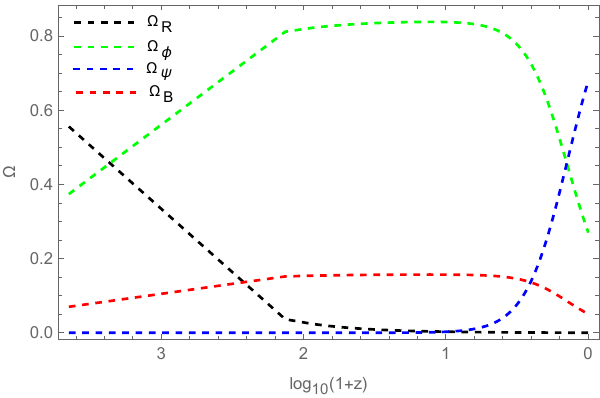}\label{nonminimal_radDE2}}
\hspace{0.2cm}
\subfigure[Evolution of density parameters ($\Omega$) with redshift (z) in a cosmological model with dark energy modeled as a phantom scalar field ($\psi$).]
{\includegraphics[width=85mm,height=50mm]{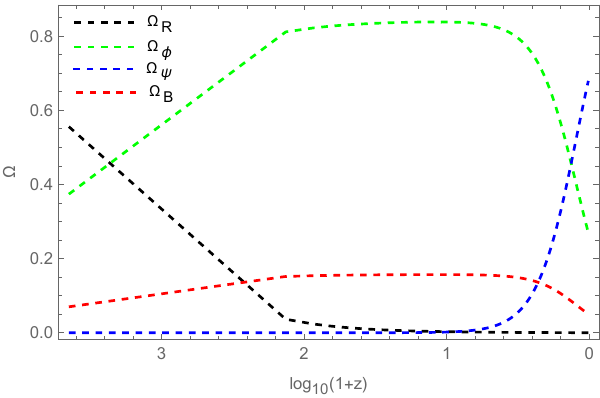}\label{nonminimal_radDE3}}
\hspace{0.2cm}
\subfigure[Evolution of time-averaged dark matter equation of state (EoS) parameter ($\omega$) with redshift (z) for different dark energy models.]
{\includegraphics[width=85mm,height=50mm]{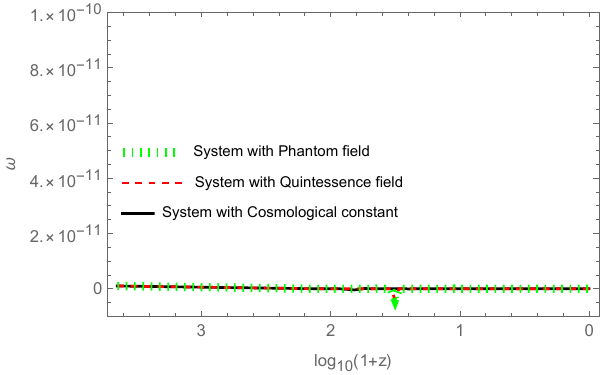}\label{nonminimal_radDE4}}
\hspace{0.2cm}
\subfigure[Evolution of dark energy equation of state (EoS) parameter ($\omega_{\text{DE}}$) with redshift (z) for different dark energy models.]
{\includegraphics[width=85mm,height=50mm]{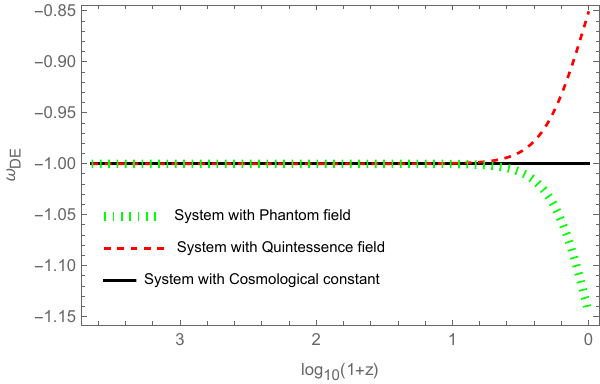}\label{nonminimal_radDE5}}
\hspace{0.2cm}
\subfigure[Evolution of effective equation-of-state (EoS) parameter $\omega_{\text{eff}} = (\langle \rho \rangle + \tilde{\rho}) / (\langle P \rangle + \tilde{P})$ with redshift for different dark energy models. Dark matter quantities are time-averaged ($\langle \cdot \rangle$), while dark energy quantities are not.]
{\includegraphics[width=85mm,height=50mm]{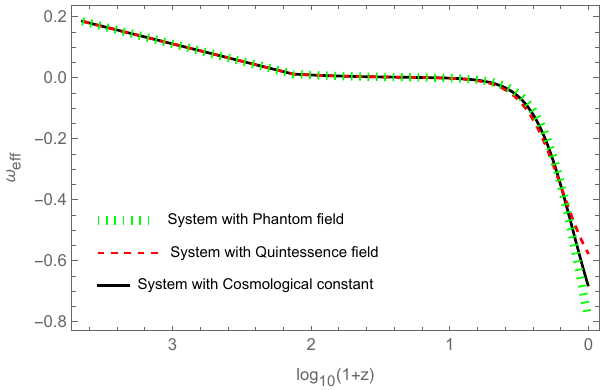}\label{nonminimal_radDE6}}
\hspace{0.2cm}
\renewcommand{\figurename}{\footnotesize Figure}
\caption{\footnotesize Evolution of cosmological variables as a function of redshift (z) in a universe composed of radiation (R), baryons (B), dark matter ($\phi$), and dark energy, where dark matter decays into radiation with decay constant $\Gamma=10^{-4}~u$. The horizontal axis shows $\log_{10}(1+z)$, plotted in decreasing order to represent the forward progression of cosmic time. Dark energy is modeled in three ways: (1) a cosmological constant with $\Lambda = 1.08\, u^2$; (2) a quintessence scalar field ($\psi$) with exponential potential $V(\psi) = V_m e^{-\lambda \psi}$, $V_m = 1.38\, u^2$, $\lambda=1$, and initial conditions $\psi(t_i) = 0.0001$, $\dot{\psi}(t_i) = 0.0001~u$; (3) a phantom scalar field with the same potential form and same initial and parameter values except $V_m = 0.91\, u^2$. Dark matter is an oscillating scalar field $\phi$ with mass $m = 1.93\times 10^{11}\, u$ and initial conditions $\phi_\pm(t_i) = 10^{-6}$, $\dot{\phi}_\pm(t_i) = -\frac{3H_i}{2} \times 10^{-6}~u$, with $a_i = 2.24\times10^{-4}$ and $H_i = 1.82\times10^5~u$. All other cosmological parameters follow Planck 2018 \cite{Planck:2018vyg}, and quantities are expressed in geometrized units $u = 10^{-28}~\mathrm{cm}^{-1}$.}
\label{nonminimal_radDE}
\end{figure*}

Here, we present the evolution of the density parameters of the individual components—radiation, baryonic matter, dark matter, and dark energy—as well as the equation-of-state (EoS) parameters for dark matter, dark energy, and the total cosmic fluid, for the case where dark matter decays into radiation with a small decay constant $\Gamma = 10^{-4}\,u$, expressed in natural units as $10^{-37}\,\mathrm{eV}$. The plotted quantities in Fig.~(\ref{nonminimal_radDE}) demonstrate that their evolution throughout cosmic history remains essentially identical to the case without dark matter decay, with all variables following the same overall trends. These results confirm that a small coupling between dark matter and radiation, corresponding to the chosen decay constant, is cosmologically viable and does not significantly alter the background evolution of the universe. The present-day values of the density parameters remain consistent with observational constraints, indicating that even such a small decay rate is compatible with standard cosmology.

\subsection{Numerical Energy-Balance Diagnostic}\label{Radiation_continuity_appa}

\begin{figure*}
\centering
\subfigure[Residual of the radiation continuity equation ($\mathcal{R}$) with redshift (z).]
{\includegraphics[width=85mm,height=50mm]{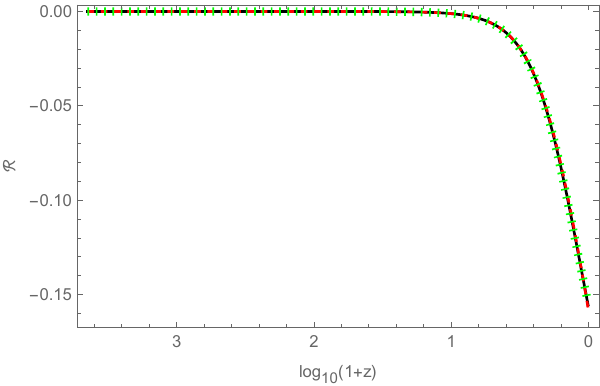}\label{radiation_continuity1}}
\hspace{0.2cm}
\subfigure[Relative percentage residual of the radiation continuity equation with redshift (z)]
{\includegraphics[width=85mm,height=50mm]{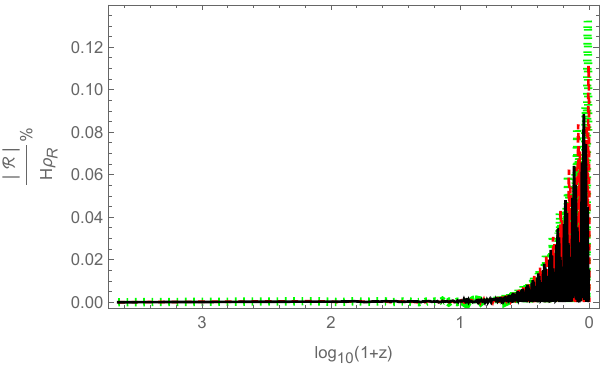}\label{radiation_continuity2}}
\hspace{0.2cm}
\renewcommand{\figurename}{\footnotesize Figure}
\caption{\footnotesize Evolution of the residual of the radiation continuity equation ($\mathcal{R}$) in $u^{3}$ units and its normalized form expressed as a percentage relative to $H\rho_{R}$, shown as functions of redshift in a universe composed of radiation (R), baryonic matter (B), dark matter ($\phi$), and dark energy. The dark matter component decays into radiation with a decay constant $\Gamma = 10^{-4}\,u$. The horizontal axis shows $\log_{10}(1 + z)$, plotted in decreasing order to represent the forward progression of cosmic time. The dark energy sector is modeled using three approaches: (1) a cosmological constant (black curve) with $\Lambda = 1.08\,u^2$; (2) a quintessence-like scalar field (red curve, $\psi$) with an exponential potential $V(\psi) = V_{m} e^{-\lambda \psi}$ and $V_{m} = 1.38\,u^2$; and (3) a phantom-like scalar field (green curve, $\psi$) with the same potential form but $V_{m} = 0.91\,u^2$. In both scalar field models, the parameter $\lambda = 1$ and initial conditions $\psi(t_i) = 0.0001$, $\dot{\psi}(t_i) = 0.0001~u$;. The dark matter field $\phi$ is modeled as an oscillating scalar field with mass $m = 1.93 \times 10^{11}\,u$ and initial conditions $\phi_\pm(t_i) = 10^{-6}$, $\dot{\phi}_\pm(t_i) = -\frac{3H_i}{2} \times 10^{-6}~u$, with $a_i = 2.24\times10^{-4}$ and $H_i = 1.82\times10^5~u$. All other cosmological parameters follow Planck 2018 \cite{Planck:2018vyg}, and quantities are expressed in geometrized units $u = 10^{-28}~\mathrm{cm}^{-1}$.}
\label{radiation_continuity}
\end{figure*}

To test the numerical accuracy of the decay runs, we have computed the residual of the radiation continuity equation, $\mathcal{R}=\dot{\rho}_{R} + 4H\rho_{R} - \Gamma\langle\dot{\phi}^2\rangle,$ along with its normalized form expressed as a percentage relative to $H\rho_R$, where the percentage uses the absolute value of $\mathcal{R}$ to quantify the maximum fractional deviation and $\rho_{R}$ is the total radiation energy density. The resulting quantities, shown in Fig. (\ref{radiation_continuity}) for the cosmological constant, quintessence, and phantom field cases, remain consistent with numerical precision throughout the evolution. The residual $\mathcal{R}$ remains close to zero throughout the evolution with respect to redshift, in agreement with the theoretical expectation. The maximum relative residual is about $0.08\%$ for the cosmological constant model, $0.10\%$ for the quintessence like scalar field dark energy and $0.12\%$ for phantom like dark energy case, demonstrating that the total energy exchange between the scalar field and radiation is well conserved by the numerical integration scheme.

\FloatBarrier


\end{document}